\pdfoutput=1 
\documentclass[]{cas-sc}
\usepackage{float}
\usepackage[numbers, sort&compress]{natbib}
\PassOptionsToPackage{hyphens}{url}
\usepackage{breakurl}
\makeatletter
\def\UrlAlphabet{%
      \do\a\do\b\do\c\do\d\do\e\do\f\do\g\do\h\do\i\do\j%
      \do\k\do\l\do\m\do\n\do\o\do\p\do\q\do\r\do\s\do\t%
      \do\u\do\v\do\w\do\x\do\y\do\z\do\A\do\B\do\C\do\D%
      \do\E\do\F\do\G\do\H\do\I\do\J\do\K\do\L\do\M\do\N%
      \do\O\do\P\do\Q\do\R\do\S\do\T\do\U\do\V\do\W\do\X%
      \do\Y\do\Z}
\def\UrlDigits{\do\1\do\2\do\3\do\4\do\5\do\6\do\7\do\8\do\9\do\0}
\g@addto@macro{\UrlBreaks}{\UrlOrds}
\g@addto@macro{\UrlBreaks}{\UrlAlphabet}
\g@addto@macro{\UrlBreaks}{\UrlDigits}
\makeatother
\def\tsc#1{\csdef{#1}{\textsc{\lowercase{#1}}\xspace}}
\tsc{WGM}
\tsc{QE}
\tsc{EP}
\tsc{PMS}
\tsc{BEC}
\tsc{DE}

\begin{document}
\let\WriteBookmarks\relax
\def\floatpagepagefraction{1}
\def\textpagefraction{.001}

\newcommand{\jx}[1] {\textcolor{blue}{#1}}

\shorttitle{Classifier-guided neural blind deconvolution}

\shortauthors{Jing-Xiao Liao et~al.}

\title [mode = title]{Classifier-guided neural blind deconvolution: a physics-informed denoising module for bearing fault diagnosis under heavy noise}     




%
\author[1,2]{Jing-Xiao Liao}





\credit{Writing - Original Draft, Methodology, Software, Data curation}

\address[1]{Department of Industrial and Systems Engineering, The Hong Kong Polytechnic University, Hong Kong, Special Administrative Region of China}

\address[2]{School of Instrumentation Science and Engineering, Harbin Institute of Technology, Harbin, China}

\address[3]{School of Mechanical, Electronic and Control Engineering, Beijing Jiaotong University, Beijing, China}



\author[3]{Chao He}
\credit{Writing - Review \& Editing, Investigation, Formal analysis, Validation}

\author[1]{Jipu Li}
\credit{Writing - Review \& Editing, Investigation, Formal analysis, Validation}



\author%
[2]
{Jinwei Sun}

\credit{Project administration, Conceptualization, Writing - Review \& Editing}




\author%
[2]
{Shiping Zhang}[type=editor,
                orcid=0000-0001-9329-8894]
\cormark[1]
\credit{Funding acquisition, Project administration, Supervision}
\cortext[cor1]{Corresponding authors}
\ead{spzhang@hit.edu.cn}

\author%
[1]
{Xiaoge Zhang}[type=editor,
                orcid=0000-0001-6831-3175]
\cormark[1]
\credit{Funding acquisition, Project administration, Supervision}
\ead{xiaoge.zhang@polyu.edu.hk}


\begin{abstract}
Blind deconvolution (BD) has been demonstrated as an efficacious approach for extracting bearing fault-specific features from vibration signals under strong background noise. Despite BD's desirable feature in adaptability and mathematical interpretability, a significant challenge persists: \textit{How to effectively integrate BD with fault-diagnosing classifiers?} This issue arises because the traditional BD method is solely designed for feature extraction with its own optimizer and objective function. When BD is combined with downstream deep learning classifiers, the different learning objectives will be in conflict. To address this problem, this paper introduces classifier-guided BD (ClassBD) for joint learning of BD-based feature extraction and deep learning-based fault classification. Firstly, we present a time and frequency neural BD that employs neural networks to implement conventional BD, thereby facilitating the  seamless integration of BD and the deep learning classifier for co-optimization of model parameters. Specifically, the neural BD incorporates two filters: i) a time domain quadratic filter to utilize quadratic convolutional networks for extracting periodic impulses; ii) a frequency domain linear filter composed of a fully-connected neural network to amplify discrete frequency components. Subsequently, we develop a unified framework to use a deep learning classifier to guide the learning of BD filters. In addition, we devise a physics-informed loss function composed of kurtosis, $l_2/l_4$ norm, and a cross-entropy loss to jointly optimize the BD filters and deep learning classifier. Consequently, the fault labels provide useful information to direct BD to extract features that distinguish classes amidst strong noise. To the best of our knowledge, this is the first of its kind that BD is successfully applied to bearing fault diagnosis. Experimental results from three datasets demonstrate that ClassBD outperforms other state-of-the-art methods under noisy conditions. We have shared our code at \url{https://github.com/asdvfghg/ClassBD}.

\end{abstract}




\maketitle

\section{Introduction}
Rotating machinery, such as aero engines, pumps, and wind turbines, plays an indispensable role in various industrial applications. However, the components that support the rotation, particularly rolling bearings, are susceptible to damage due to long working hours in high temperature, high speed, and other harsh conditions~\cite{randall2011industrial, miao2022review}. {The damage of bearings, i.e., cage fracture and race crack, causes unexpected machinery failures and leads to costly downtime and even catastrophic outcomes.} Therefore, timely and accurate diagnosis of bearing faults is of great importance for ensuring the sound and reliable operations of rotating machinery~\cite{wang2022bearing}. Nevertheless, one of the major challenges in bearing fault diagnosis is that the measured vibration signals are often contaminated by background noise arising from complex transmission paths {(mechanical transmission systems)} and environmental sources {(coupled vibration source from multiple machines)}. These noises substantially obscure and distort key information that is important for discriminating faults in rotating machinery. Hence, developing effective methods for extracting fault-specific features from noisy signals has emerged as an active research topic.

Presently, methodologies for signal denoising fall into two distinct categories: data-driven methods and signal processing methods. Despite the significant advances made by data-driven methods in recent years, these models fall short in transparency, which hinders their utility in decision-making processes~\cite{li2023ifd, li2024dconformer, zhang2020deep, zuo2023hybrid, liang2024multi}. On the other hand, a plethora of signal processing approaches with rigorous mathematical foundations have been proposed for extracting fault-related features. These include, but are not limited to, wavelet transform (WT)~\cite{lou2004bearing, peng2005comparison, kankar2011rolling, xu2019novel}, variational mode decomposition (VMD)~\cite{2013variational, wang2015research}, singular value decomposition (SVD)~\cite{stewart1993early, li2020bearing}, spectral kurtosis (SK)~\cite{vrabie2003spectral, antoni2006spectral, wang2016spectral}, cyclostationary analysis~\cite{mccormick1998cyclostationarity, mcdonald2012maximum}, and blind deconvolution (BD)~\cite{miao2022review, cabrelli1985minimum, buzzoni2018blind}. Among them, the BD method possesses some unique advantages due to its adaptability and lack of constraints on bandwidth or center frequency {(filter's maximum gain frequency)}, and these features make BD an ideal tool for extracting repeated transient impulses~\cite{he2021optimized}. \jx{} Hence, we concentrate on the exploring the BD-based method for fault diagnosis in this paper.



In essence, the BD method recovers the input signal features from the output signals when both the system and the input signals are unknown~\cite{Haykin}. {In other words, BD only uses the measured signal to reconstruct the fault source signal by estimating the transmission path function.} When processing vibration signal, BD optimizes an adaptive finite impulse response (FIR) filter to recover the repeated transient impulses, which are regarded as the informative features for bearing fault classification~\cite{miao2022review}. {A key issue in BD is how to design an objective function to effectively characterize the fault impulsive signatures, such as sparsity and cyclic periodicity.}
For example,~\citet{wiggins1978minimum} proposed the first of its kind in BD coined as the minimum entropy deconvolution (MED) in 1978 for {non-stationary signal denoising.} MED used kurtosis~\cite{pearson1905fehlergesetz} as the objective function to search for an optimal inverse filter. However, kurtosis is only sensitive to outliers, and it thus fails to distinguish between random impulses and cyclic impulses~\cite{cheng2019adaptive}. As a result, many other objective functions are subsequently developed to capture cyclic information, such as the maximum correlated kurtosis deconvolution (MCKD)~\cite{mcdonald2012maximum}, the multipoint optimal MED adjusted (MOMEDA)~\cite{mcdonald2017multipoint}, the second-order cyclostationarity blind deconvolution (CYCBD)~\cite{buzzoni2018blind}, and the adaptive cyclostationarity blind deconvolution (ACYCBD)~\cite{wang2022bearing}. These methods share a common goal in attempt to address the extraction of bearing fault-related characteristics from vibration signals in the presence of complex noise.

However, a fundamental challenge of applying BD to bearing fault diagnosis remains unresolved: \textit{How to effectively integrate BD with fault diagnosing classifiers?} Existing BD methods often assess BD performance by examining the recovered signals, but only a few works have attempted to integrate BD and convolutional neural networks (CNN) for {end-to-end} fault diagnosis~\cite{wang2020minimum}. {Combination of BD and classifiers fails to yield optimal performance and even exerts a detrimental effect on the classification task. The main reason is that BD and the classifier operate in two separate optimization spaces, which possess distinct optimizers (filter's optimizer vs CNN's optimizer) and divergent optimization objectives (BD objective function vs cross-entropy loss function). This leads to a lack of consistency during training.}
For instance, BD may enhance the cyclic impulse of the fault signal, but at the same time, it may diminish the differences between various fault severities. Therefore, there is a need for a unified framework that can coherently and efficiently integrate BD and classifiers.

In this paper, we propose a novel framework that uses neural networks to perform both BD and classification. First, we employ a neural BD to process the raw vibration signal. Established upon the multi-task neural network blind deconvolution (MNNBD)~\cite{Liao_2023}, we replace the conventional BD filters with neural networks. The neural BD gives rise to two advantages: i) It implement multi-channel and multi-layer filters by using convolutional kernels as adaptive filters, while the conventional BD filters are usually single-channel; ii) It employs the optimizer of convolutional neural networks (CNNs) to find the optimal filter coefficients, while the conventional BD methods rely on less-efficient matrix operations \cite{wiggins1978minimum, buzzoni2018blind} or particle swarm optimization (PSO) \cite{cheng2018application, cheng2019novel}. Moreover, such neural BD can be easily integrated with deep learning classifiers to achieve co-optimization {of weight parameters}. 

{Our proposed BD framework includes two neural network modules: a time domain quadratic convolutional filter and a frequency domain linear filter (we use the term “filter” to remain consistent with the terminology in the field of signal processing and BD). The former with two layers of symmetric quadratic convolutional neural networks (QCNN)~\cite{fan2018new, 10076833} excels in extracting periodic impulses in the time domain. The latter composed of a fully-connected neural network filters signals in the frequency domain post-Fast Fourier transform (FFT), thus enhancing the capability to filter the signal’s frequency components.} {Furthermore, inspired by advances in physics-informed neural networks~\cite{he2023idsn, ni2023physics, yang2024physics}, we introduce a unified framework ClassBD to integrate BD and deep learning classifiers. ClassBD transforms conventional BD, typically an unsupervised learning problem, into an supervised learning task using fault labels. This guides BD in extracting class-distinguishing features amidst noise. Our threefold pipeline includes neural BD as a plug-and-play module in the first layer of deep learning classifier, a physics-informed loss function optimizing both BD filters and classifiers, and an uncertainty-aware weighing loss strategy balancing the three loss components during training.} Our contributions are summarized as follows:

\begin{enumerate}
    \item We introduce a plug-and-play time and frequency neural blind deconvolution module. This module comprises two cascaded components: a quadratic convolutional neural filter and a frequency linear neural filter. From a mathematical perspective, we demonstrate that the quadratic neural filter enhances the filter’s capacity to extract periodic impulses in the time domain. The linear neural filter, on the other hand, offers the ability to filter signals in the frequency domain and it leads to a crucial enhancement for improving BD performance.

    \item We develop a unified framework -- ClassBD -- to integrate BD and deep learning classifiers. By employing a deep learning classifier to guide the learning of BD filters, we transition from the conventional unsupervised BD optimization to supervised learning. The fault labels supply useful information in guiding the BD to extract class-distinguishing features amidst background noise. To the best of our knowledge, this is the first BD method of its kind to achieve bearing fault diagnosis under heavy noise while providing good interpretability.
\end{enumerate}

The rest of the paper is organized as follows. Section 2 introduces some background knowledge of blind deconvolution in signal processing and quadratic neural networks. Section 3 is the proposed method in detail. In Section 4, we conduct computational experiments in two public and one private datasets. Section 5 analyzes the properties of the proposed method. Section 6 is the conclusions.

\section{Preliminaries}
{Table~\ref{tab:symbols} presents some important symbols that will be used later in this paper.}

\begin{table}[pos=h]
\centering
\caption{A list of mathematical notations and symbols.}
\begin{tabular}{@{}ll@{}}
\toprule
Symbol                   & Description                                        \\ \midrule
$N$                     & Length of the input, constant \\
$K$                     & Length of the filter (convolution kernel), constant ($K < N$)\\
$\boldsymbol{x} \in \mathbb{R}^{1\times N}$ \& $x(n), \ n=1,\cdots,N$                       & Input signal measured by sensors                 \\
$\boldsymbol{d} \in \mathbb{R}^{1\times N}$                       & Fault source signal                                \\
$\boldsymbol{n} \in \mathbb{R}^{1\times N} \& \  n(n), \ n=1,\cdots,N$                     & Additive noise                                     \\
$\boldsymbol{h}_d \in \mathbb{R}^{1\times K}$                       & Transfer function in accordance to fault  source                   \\
$\boldsymbol{h}_n \in \mathbb{R}^{1\times K}$                       & Transfer function in accordance to noise                         \\
$\boldsymbol{f} \in \mathbb{R}^{1\times K}$                        & Blind deconvolution filter                        \\
$\boldsymbol{y} \in \mathbb{R}^{1\times N} \& \  y(n), \ n=1,\cdots,N$                        & Recovered signal after blind deconvolution                                \\
$\mathcal{K}(\cdot)$                        & Blind deconvolution objective function    \\
$\boldsymbol{W} \in \mathbb{R}^{1\times K} \&  \  w(n), \ n=1,\cdots,K$     & Weight parameters of neural network                \\
$b$                        & Bias parameters of neural network                \\
$p(\cdot)$                      & Impulse response of bearing fault                     \\
$q(\cdot)$                     & Periodic modulation signal                         \\
$T$                     & Cyclic period, constant                         \\
$R_{xx}$                        & Instantaneous autocorrelation function              \\
$\hat{x}(n), \ n=1,\cdots,N$ & Signal after time domain filter                    \\
$\mathcal{F}(\cdot), \mathcal{F}^{-1}(\cdot)$                        & Fast Fourier Transform (FFT) and Inverse Fast Fourier Transform (IFFT)                          \\
$\hat{X}(f), \ f=1,\cdots,N$                         & Signal after FFT (frequency domain)                \\
$\hat{Y}(f), \ f=1,\cdots,N$                        & Signal after frequency domain filter (frequency domain) \\
$\hat{y}(n), \ n=1,\cdots,N$                        & Signal after IFFT (time domain)                      \\
$h(n),  \ n=1,\cdots,N$                        & Hilbert Transform (HT) of the signal in time domain  \\
$H(f), \ f=1,\cdots,N$                        & Hilbert Transform of the signal in frequency domain  \\
$z(n), \ n=1,\cdots,N$                        & Analytic signal (time domain) \\
$ES(f),  \ f=1,\cdots,N$                       & Envelope spectrum (frequency domain)                         \\
$\mathcal{L}_c, \mathcal{L}_t, \mathcal{L}_f$                       & Cross-entropy, $l_4/l_2$ norm and $l_2/l_4$ norm loss functions                                  \\
$\mathcal{G}_{l_p/l_q}(\cdot)$                        & \multicolumn{1}{l}{Generalized sparsity blind deconvolution objective function} \\ \bottomrule
\end{tabular}
\label{tab:symbols}
\end{table}

\subsection{Blind deconvolution}

In the realm of non-stationary signal processing, deconvolution reverses the effects of convolution operations performed by a linear time-invariant system on the input signal. A specialized form of this technique, known as blind deconvolution (BD) or more accurately, unsupervised deconvolution, aims to utilize the output signal to recover the input signal when both the {signal transfer system} and the input signal are unknown~\cite{Haykin}. In the context of vibration signals associated with rotating machinery, the measured signals can be interpreted as the result of a convolution operation between the cyclic fault impulses and the transfer path functions originating from the fault source to the sensor~\cite{miao2022review}. From a mathematical perspective, given the measured signal $\boldsymbol{x} \in \mathbb{R}^N$, the fault source signal $\boldsymbol{d} \in \mathbb{R}^N$, and the additive noise $\boldsymbol{n} \in \mathbb{R}^N$, the signal transfer process can be defined as follows:
\begin{equation}
\boldsymbol{x} = \boldsymbol{d} * \boldsymbol{h}_d + \boldsymbol{n} * \boldsymbol{h}_n,
\label{eq:convolution}
\end{equation}
where $\boldsymbol{h}_d$ and $\boldsymbol{h}_n$ represent the transfer functions, and $*$ denotes the convolution operation. Hereafter, bold-faced symbols will be used to denote matrices.

The objective of BD is to extract fault-related features (cyclic impulses) from the measured signal. Towards this goal, it aims to recover the signal $\boldsymbol{y} \in \mathbb{R}^{N}$ that is closer to the fault source by  constructing a filter $\boldsymbol{f} \in \mathbb{R}^{L}$:
\begin{equation}
\boldsymbol{y} = \boldsymbol{x} * \boldsymbol{f} \approx \boldsymbol{d}.
\end{equation}

However, due to the complexity of machinery systems, it is often impractical to accurately estimate the transfer function and its frequency response. This challenge is further compounded by the presence of unpredictable noise. Consequently, in the absence of prior information, {such as an accurate fault impulse period}, BD is considered as an ill-posed problem. Given the non-stationary and periodic nature of the fault characteristics, a variety of sparsity indexes have been proposed to function as optimization objective functions~\cite{he2021optimized, he2021extracting, buzzoni2018blind, mcdonald2017multipoint}. A representative example is kurtosis~\cite{pearson1905fehlergesetz}, which is utilized as the objective function in MED~\cite{wiggins1978minimum}:
\begin{equation}
\mathcal{K}= \frac{\sum_{n=1}^{N}{y(n)^4}}{(\sum_{n=1}^{N}{y(n)^2)^2}}.
\label{eq:kurt}
\end{equation}
{where $y(n)$ denotes the output of the BD filter, and its length is equal to the input $N$.}

Essentially, Kurtosis is a statistical quantity that assesses the data distribution. An increase in the kurtosis value indicates a deviation from the standard normal distribution~\cite{pearson1905fehlergesetz}. Intuitively, cyclic impulses emerge in the vibration signals when the fault occurs, and the kurtosis value of vibration signal gets increased due to the presence of more peaks (outliers). Consequently, maximizing kurtosis drives the adaptive filter to recover more impulses. Naturally, the optimization objective is defined as follows:
\begin{equation}
\begin{aligned}
&\max_{\boldsymbol{f}}\mathcal{K}(\boldsymbol{y}) \\
&s.t. \quad \boldsymbol{y} = \boldsymbol{x} * \boldsymbol{f}, \ \|\boldsymbol{f}\|_{l_2}=1.
\end{aligned}
\label{eq:opt}
\end{equation}

Several effective optimization methods have been developed for BD, including matrix operations \cite{wiggins1978minimum, buzzoni2018blind}, particle swarm optimization (PSO) \cite{cheng2018application, cheng2019novel}, and backpropagation \cite{fang2021blind, fang2022minimum}. Obviously, the performance of BD is strongly influenced by the choice of optimization method.In recent years, considerable efforts have been made to identify more general objective functions, design new filters, and devise more powerful optimization tools~\cite{miao2022review}.

\subsection{Quadratic neural networks}

The concept of high-order neural networks, also known as polynomial neural networks, has its roots in the 1970s. The Group Method of Data Handling (GMDH), which utilizes a polynomial network as a feature extractor, was first proposed by~\citet{ivakhnenko1971polynomial}. Subsequently,~\citet{shin1991pi} introduced the pi-sigma network to incorporate high-order polynomial operators: 
\begin{equation}
    y_i = \sigma(\prod_{j}({\sum_{k}{w_{kji}{x}_k+b_{ji}}})).
\end{equation}
{where $w$, $b$ are learnable parameters, $\sigma(\cdot)$ represents the activation function, and $x$ denotes the input. The high-order polynomials are implemented by multiplying several linear functions.}


In recent years, advances in deep learning have provided a platform for re-examination and integration of polynomial operators into fully-connected neural networks and convolutional neural networks. Methodologies for introducing polynomials into neural networks can be categorized into two classes: polynomial structure and polynomial neuron. In terms of the former, polynomial neural networks were developed using polynomial expansion via recursion~\cite{chrysos2023regularization} or tensor decomposition~\cite{chrysos2021deep, chrysos2022augmenting}. For the latter, the linear function (neuron) in traditional neural networks was substituted with various polynomial functions~\cite{fan2018new,xu2022quadralib}. This study primarily focuses on the neuron-level methods.

While polynomial functions can be extended to higher orders, this significantly increases the computational complexity of the neural network. To facilitate stable training, the polynomial function is typically restricted to the second order, which is known as quadratic neural network.

Mathematically, given the input $\boldsymbol{x} \in \mathbb{R}^{d}$, the one-layer traditional neural network can be expressed as: 
\begin{equation}
    y=\sigma(\boldsymbol{W}^\top \boldsymbol{x} +b),
\label{eq:linear}
\end{equation}
where $\boldsymbol{W}^\top \in \mathbb{R}^{d}$ and $b$ indicate weight and bias of neural network, and $\sigma(\cdot)$ is the activation function. 

A quadratic network is constructed by replacing the linear function with a quadratic function~\cite{fan2018new}:
\begin{equation} 
y=\sigma((\boldsymbol{W}_1^{\top}\boldsymbol{x}+b_1)(\boldsymbol{W}_2^{\top}\boldsymbol{x}+b_2)+\boldsymbol{W}_3^{\top}(\boldsymbol{x}\odot\boldsymbol{x})+b_3),
\label{eq:quadratic}
\end{equation}
where $\odot$ is Hadamard product. 

It is noteworthy that a variety of quadratic neurons have been proposed in the literature as summarized in Table~\ref{tab:neurons}. In this paper, we opt for the version proposed by~\citet{fan2018new} (see Eq.~\eqref{eq:quadratic}). {Compared the quadratic neuron proposed by Fan \textit{et al.} and others (Bu\&Karpatne, Xu \textit{et al.)}, it serves as a general version which consists of entire inner-product term and power term.}


\begin{table}[!ht]
\centering
\caption{A summary on quadratic neurons in neural network, where $\boldsymbol{W} \in \mathbb{R}^{n\times n}$, $\boldsymbol{W}_i \in \mathbb{R}^{n\times 1}$, and the bias terms are omitted.}
\begin{tabular}{ll}
\toprule
Authors           & Quadratic functions          \\ \midrule
Zoumpourlis \textit{et al.}(2017) \cite{zoumpourlis2017non, micikevicius2017mixed} & $y=\sigma(\boldsymbol{x}^{\top}\boldsymbol{W}\boldsymbol{x}+\boldsymbol{W}^\top\boldsymbol{x})$               \\ \midrule
Jiang \textit{et al.}(2019) \cite{jiang2020nonlinear};      & \multirow{2}{*}{$y=\sigma(\boldsymbol{x}^{\top}\boldsymbol{W}\boldsymbol{x}$)} \\ 
Mantini\&Shah(2021) \cite{mantini2021cqnn}     &                        \\ \midrule
Goyal \textit{et al.}(2020) \cite{goyal2020improved}    & $y=\sigma(\boldsymbol{W}^\top(\boldsymbol{x}\odot\boldsymbol{x}))$               \\ \midrule
Bu\&Karpatne(2021) \cite{bu2021quadratic}       & $y=\sigma((\boldsymbol{W}_1^\top\boldsymbol{x})(\boldsymbol{W}_2^\top\boldsymbol{x}))$             \\ \midrule
Xu \textit{et al.}(2022) \cite{xu2022quadralib} & $y=\sigma((\boldsymbol{W}_1^\top\boldsymbol{x}) (\boldsymbol{W}_2^\top\boldsymbol{x})+\boldsymbol{W}_3^\top\boldsymbol{x})$ \\
\midrule
Fan \textit{et al.}(2018)   \cite{fan2018new}      & $y=\sigma((\boldsymbol{W}_1^\top\boldsymbol{x})(\boldsymbol{W}_2^\top\boldsymbol{x})+\boldsymbol{W}_3^\top(\boldsymbol{x}\odot\boldsymbol{x}))$        \\ \bottomrule
\end{tabular}
\label{tab:neurons}
\end{table}

\section{Methodology}

\begin{figure}[!ht]
    \centering
    \includegraphics[width=0.8\linewidth]{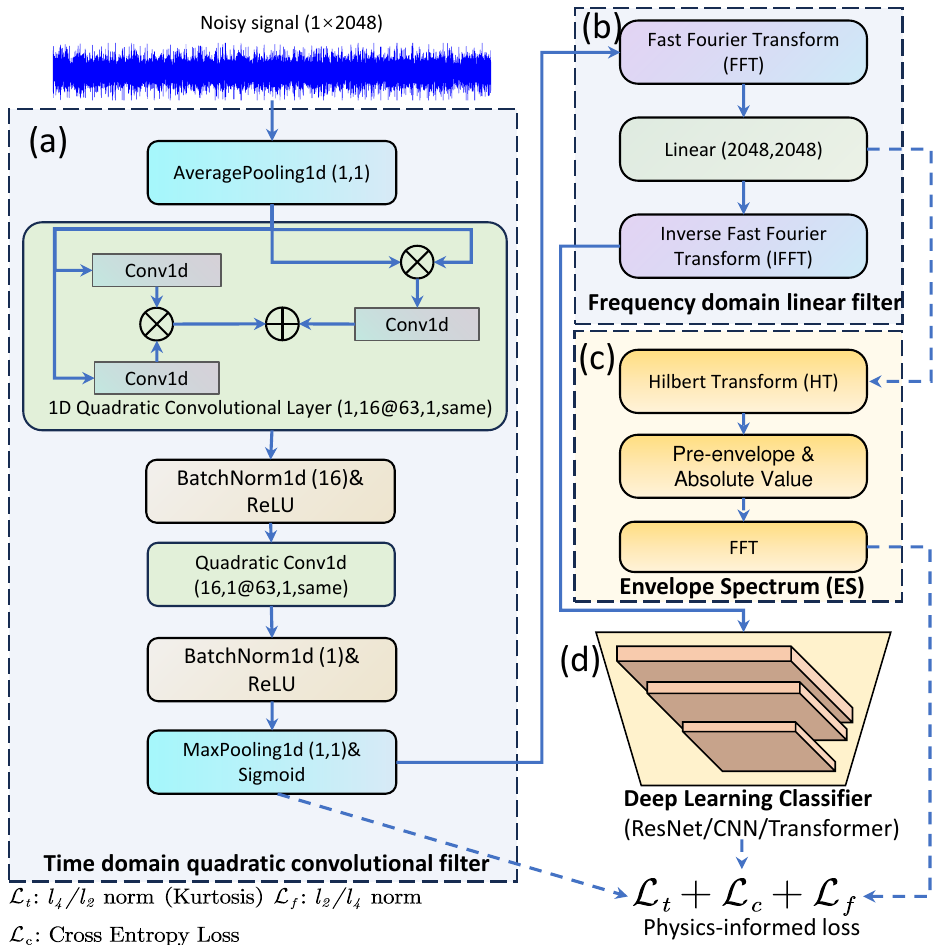}
    \caption{  
    The proposed framework: (a) The time-domain filter, consisting of two symmetric quadratic convolutional neural network (QCNN) layers, is designed for time domain BD. (b) The frequency-domain filter, composed of a fully-connected layer, is utilized for frequency domain BD. (c) The output from the fully-connected layer is extracted to compute the envelope spectrum (ES), which is crucial for constructing the objective function. (d) The output from the frequency domain linear filter is directed to the deep learning classifier to yield classification results.}
    \label{fig:structure}
\end{figure}

The proposed framework, as illustrated in Figure~\ref{fig:structure}, primarily consists of two BD filters, namely a time domain quadratic convolutional filter and a frequency domain linear filter. These filters serve as a plug-and-play denoising module, and they are designed to perform the same function as conventional BD methods to ensure that the output is in the same dimension as the input.
\begin{enumerate}
    \item The time domain filter is characterized by two symmetric quadratic convolutional neural network (QCNN) layers. A 16-channel QCNN is employed to filter the input signal (1 $\times$ 2048), and an inverse QCNN layer is used to fuse the 16 channels into one for recovering the input signal.
    
    \item The frequency domain filter, on the other hand, starts with the fast Fourier transform (FFT) with an emphasis on highlighting the discrete frequency components. Subsequently, a linear neural layer filters the frequency domain of the signals, and the inverse FFT (IFFT) is conducted to recover the time domain signal. Moreover, an objective function in the envelope spectrum (ES) is designed for optimization.
\end{enumerate}

After the neural BD filters, 1D deep learning classifiers, such as ResNet, CNN, or Transformer, can be directly used to recognize the fault type. In this paper, we employ a popular and simple network -- wide first kernel deep convolutional neural network (WDCNN)~\cite{zhang2017new} -- as our classifier. Finally, a physics-informed loss function is devised as the optimization objective to guide the learning of the model. This function comprises a cross-entropy loss $\mathcal{L}_c$ and a kurtosis $\mathcal{L}_t$, and $l_2/l_4$ norm $\mathcal{L}_f$. It should be noted that $\mathcal{L}_t$ and $\mathcal{L}_f$ are used to calculate the statistical characteristics of the outputs of the time filter and frequency filter, respectively.


\subsection{Time domain quadratic convolutional filter}
As mentioned earlier, quadratic convolutional neural networks (QCNN) is a key constitutional component in the time domain filter. In this paper, we utilize the quadratic neuron proposed by~\citet{fan2018new}, which is expressed as follows:
\begin{equation}
\boldsymbol{y}=\sigma((\boldsymbol{W}_1*\boldsymbol{x}+\boldsymbol{b}_1)\odot(\boldsymbol{W}_2*\boldsymbol{x}+\boldsymbol{b}_2)+\boldsymbol{W}_3*(\boldsymbol{x}\odot\boldsymbol{x})+\boldsymbol{b}_3).
\label{eq:cquadratic}
\end{equation}
where $*$ denotes the convolutional operation. 

{Currently, quadratic networks have been demonstrated to possess certain advantages in both theoretical and practical aspects. Firstly, in terms of efficiency, quadratic networks are capable of using neurons at a polynomial level to approximate radial functions, while conventional neural network necessitate neurons at an exponential level~\cite{fan2023one}. Secondly, when it comes to feature representation, quadratic networks can achieve polynomial approximation. In contrast, conventional networks resort to piece-wise approximation via non-linear activation functions. The polynomial approximation is better than the piece-wise one in representing complex functions~\cite{fan2020universal}. Lastly, in practical applications, several studies have successfully incorporated quadratic neural networks into bearing fault diagnosis, and they have reported superior performance under challenging conditions such as strong noise~\cite{10076833}, data imbalance~\cite{yu2023classweighted}, and variation loads~\cite{10462123}. This further underscores the practical utility and robustness of quadratic networks in fault diagnosis.}


{Despite the impressive performance of quadratic networks, there is a substantial increase in the number of model parameters and non-linear multiplication operations in quadratic networks. Accordingly, the number of parameters to be optimized has increased largely.} Previous studies have shown that conventional initialization techniques can significantly hinder the convergence of quadratic networks~\cite{fan2023expressivity, chrysos2023regularization}. To overcome this problem, we design a dedicated strategy to initialize the quadratic network:
\begin{equation}
\begin{aligned}
    \mathrm{Normal} \ \mathrm{initialization}&\left\{ \begin{array}{c}
	\boldsymbol{W}_1\sim \mathcal{N} (0,\sigma ^2), \ \mathrm{with} \ \sigma =\sqrt{1/32(k_n)},\\
	\boldsymbol{b}_1\sim \,\,\mathcal{U} \left( -B,B \right), \ \mathrm{with} \ B =\sqrt{1/k_n}.\\
\end{array} \right.
\\
\mathrm{Zero} \ \mathrm{initialization}&\begin{cases}
	\boldsymbol{W}_2=\boldsymbol{0},\boldsymbol{W}_3=\boldsymbol{0},\\
	\boldsymbol{b}_2=\boldsymbol{1},b_3=\boldsymbol{0}.\\
\end{cases}
\end{aligned}
\end{equation}
where $\mathcal{N}(0,\sigma ^2)$ represents a Gaussian distribution with zero mean, $\mathcal{U} \left( -B,B \right)$ represents an uniform distribution within the range $(-B, B)$, and $k_n$ denotes the kernel size of $\boldsymbol{W}_1$. 

The group initialization strategy, also known as ReLinear~\cite{fan2023expressivity}, aims to compel QCNN to commence from an approximately first-order linear neuron. {The initial values of high-order weights are set to zero so that it grows slowly. This strategy greatly increases the stability of quadratic networks during training by avoiding gradient explosion.} In terms of implementation, we employ two QCNN layers to form a symmetric structure {which mimics a multi-layer deconvolution filter.} The first QCNN layer maps the input into 16 channels, while the second one consolidates these 16 channels into a single output. The dimension of the output is deliberately maintained the same as the input. This operation effectively implements a conventional BD filter using a convolutional neural network. Finally, as the QCNN functions as a time-domain BD, the widely-used time-domain BD objective function kurtosis (Eq.~\eqref{eq:kurt}) is employed in this filter. Thus, we construct the time domain BD loss as follows:
\begin{equation}
\mathcal{L}_t = -\frac{\sum_{n=1}^{N}{y(n)^4}}{(\sum_{n=1}^{N}{y(n)^2)^2}}.
\label{eq:lt}
\end{equation}


\subsection{Superiority of cyclic features extraction by QCNN}
In this section, we provide a theoretical derivation to address the following question: \textit{Why are quadratic convolutional networks beneficial for the extraction of features from periodic and non-stationary signals?}

Denote the input signal as $\boldsymbol{x}=[x(1), x(2), \cdots, x(N)]$, three different weight parameters (Fan's quadratic neuron) as $\boldsymbol{W}_i = [w_i(1),w_i(2), \cdots, w_i(K)]$, where $i=1,2,3$ and $K < N$. We can convert Eq.~\eqref{eq:quadratic} into a sum-product form for simplification (note that the bias terms are omitted for the sake of simplicity):
\begin{equation}
y(n)=[\sum_{i=1}^{K}w_1(i)x(n-i)][\sum_{i=1}^{K}w_2(i)x(n-i)] +  \sum_{i=1}^{K}w_3(i)x^2(n-i).
\label{eq:qelement}
\end{equation}

As shown in Eq.~\eqref{eq:qelement}, a quadratic network employs a convolution kernel of size $K$ to convolve over a segment of $\boldsymbol{x}$. The quadratic function has two parts: the product of two inner-product terms and one power term. The inner-products terms can be further factorized as follows:
\begin{equation}
\begin{aligned}
        &[\sum_{i=1}^{K}w_1(i)x(n-i)][\sum_{i=1}^{K}w_2(i)x(n-i)] \\
        = &[w_1(1)x(n-1)+w_1(2)x(n-2)+\cdots w_1(K)x(n-K)][w_2(1)x(n-1)+w_2(2)x(n-2)+\cdots w_2(K)x(n-K)] \\
        = &w_1(1)(\sum_{j=1}^{K}w_2(j)x(n-1)x(n-j))+ w_1(2)(\sum_{j=1}^{K}w_2(j)x(n-2)x(n-j)) + \cdots + w_K(2)(\sum_{j=1}^{K}w_2(j)x(n-K)x(n-j)) \\
        = & \sum_{i=1}^{K}\sum_{j=1}^{K}w_1(i)w_2(j)x(n-i)x(n-j) \\
        = & \sum_{i=1}^{K}\sum_{j=1}^{K}f(i,j)x(n-i)x(n-j),
\end{aligned}
\label{eq:firstterm}
\end{equation}
with $f(i,j) = w_1(j)w_2(i)$. 

Eq.~\eqref{eq:firstterm} demonstrates that the inner-product terms signifies several convolution operations: cross-correlation between filters and inputs, and autocorrelation for input signals. These operations are crucial for cancelling noise in the bearing fault vibration signals.

Next, we establish the relations between QCNN and bearing fault signals. The ideal bearing fault mathematical model is given as follows~\cite{antoni2006spectral, randall2001relationship}:
\begin{equation}
    x(t) = \sum_{i=-\infty}^{+\infty}p(t-iT)q(iT) + n(t),
\label{eq:signal}
\end{equation}
where $p(t)$ is an impulse response by signal impact, $q(t)=q(t+T)$ is the periodic modulation with a period of $T$, $T$ denotes the interval time between two consecutive impacts on the fault, and $n(t)$ denotes the additive Gaussian noise. 

The instantaneous autocorrelation function is defined as follows~\cite{antoni2007cyclic, randall2021vibration}:
\begin{equation}
    R_{xx}(t, \tau) =  \sum(x(t-\tau/2)x(t+\tau/2)),
\label{eq:autocor}
\end{equation}
where $\tau$ is the time delay. 

Generally, the bearing fault signal can be regarded as a second-order cyclostationary signal~\cite{randall2001relationship}. In other words, this signal presents periodicity in its second-order statistics (autocorrelation function). For Eq.~\eqref{eq:signal}, we have:
\begin{equation}
    R_{pp}(t, \tau) = R_{pp}(t+iT, \tau).
\label{eq:autoCS}
\end{equation}

 Under this assumption, since the noise is randomized over the full time period and it has no periodicity, its autocorrelation function is expressed as:
\begin{equation}
    R_{nn}(t, \tau) = 0, \tau \ne 0.
\end{equation}

{Lastly, let us show how the quadratic neuron enhances the fault-related signal from the noise.} Combining Eqs.~\eqref{eq:firstterm} and~\eqref{eq:signal}, we have:
\begin{equation}
\begin{aligned}
        &\sum_{i=1}^{K}\sum_{j=1}^{K}f(i,j)x(n-i)x(n-j) \\
        =&\sum_{i=1}^{K}\sum_{j=1}^{K}f(i,j)[\sum_{i=1}^{K}\sum_{j=1}^{K}(\sum_{k=-\infty}^{+\infty}p(t-i-kT)q(kT) + n(t-i))(\sum_{k=-\infty}^{+\infty}p(t-j-kT)q(kT) + n(t-j))] \\
        \overset{(i)}{=} & \sum_{i=1}^{K}\sum_{j=1}^{K}f(i,j)(R_{pp}(t,1) + 2R_{pn}(t, 1) + R_{nn}(t, 1)) \\
        \overset{(ii)}{\approx} & \sum_{i=1}^{K}\sum_{j=1}^{K}f(i,j)R_{pp}(t,1).
\end{aligned}
\label{eq:qcore}
\end{equation}
where (i) follows from the convolutional step being 1, that is $\tau = 1$; (ii) follows from $p(t)$ and $n(t)$ being irrelevance, such that $R_{pn}(t, 1) = 0$. In the ideal case, $R_{pp}$ is not equal to 0 only on the cycle period. {Clearly, the autocorrelation operation embedded in the QCNN enhances the second-order cyclostationary signal while suppressing the noise at the same time. Eq.~\eqref{eq:qcore} only keeps the fault-related signals.}


In addition, the second term in Eq.~\eqref{eq:qelement}, denoted as $\sum_{i=1}^{K}w_3(i)x^2(n-i)$, also plays a crucial role, as it calculates the power of the signal. This computation facilitates the amplification of the disparity between non-stationary impulses and stationary signals. Consequently, the QCNN is capable of enhancing the cyclostationary fault impulse while simultaneously suppressing the effect of random noise. Clearly, the conventional neural networks fall short in this regard. To showcase the superiority of QCNN, we conduct a comparative analysis of the feature extraction performance between quadratic and conventional networks in Section~\ref{sec:qvc}. Please refer to Section~\ref{sec:qvc} for a detailed analysis on the performance of features extracted by regular neural networks and quadratic convolutional operators.


\subsection{Frequency domain linear filter with envelope spectrum objective function}

{In our framework, the frequency domain filter serves as an auxiliary module, and it is endowed with the capability to directly manipulate the signal’s frequency domain. The main idea involves the utilization of the neural network as a filter within the frequency domain, facilitated by the Fourier transform. This approach is commonly referred to as Fourier neural networks~\cite{chi2020fast,NEURIPS2023_e66309ea}.} 

Denote the signal passing through the time domain filter as ${\hat{x}(t)}$,  the FFT $\mathcal{F}(\cdot)$ is applied to convert the signal into the frequency domain:
\begin{equation}
\hat{X}(f) = \mathcal{F}(\hat{x}(t)).
\end{equation}

{According to the Convolution Theorem, the convolution of two time-domain data is equivalent to the inner product in their Fourier transform domain~\footnote{\url{https://en.wikipedia.org/wiki/Convolution_theorem}}. Such that, a frequency filter employs a linear operation to filter the signal within the frequency domain, thereby substituting the convolution operation in the time domain. Consequently, it is reasonable to utilize a fully-connected neural network for the implementation of the frequency filter:}

\begin{equation}
\hat{Y}(f) = \sum_{f=1}^N w_f(f) \hat{X}(f) + b_f,
\end{equation}
where $w_f, b_f$ are the weights and biases in the frequency domain, {$\hat{Y}(f)$ denotes the filtered signal in frequency domain.}

Subsequently, the IFFT $\mathcal{F}^{-1}$ is employed to recover the signal to the time domain:
\begin{equation}
    \hat{y}(t) = \mathcal{F}^{-1}(\hat{Y}(f)).
\end{equation}

Second, an objective function for the frequency domain filter is also required. Previous works have proposed some BD objective functions for the frequency domain, {such as envelope spectra $l_1/l_2$ norm~\cite{peter2013design}, envelope spectra kurtosis (ESK)~\cite{zhang2016kurtosis}, and $l_p/l_q$ norm~\cite{he2021extracting}. The fundamental concept underlying these approaches is the enhancement of signal sparsity in the frequency domain, which effectively mitigates the impact of noisy frequency components.} We adopt this idea to design the objective function based on the envelope spectrum (ES). Mathematically, the Hilbert transform (HT) is defined as:
\begin{equation}
    h(t) = \frac{1}{\pi}\int_{-\infty}^{+\infty}x(\tau)\frac{1}{t-\tau}d\tau=x(t)*\frac{1}{\pi t}.
\end{equation}


In our study, the HT is applied after the linear layer, {which is computed in the frequency domain}. According to the Convolution Theorem, we have:
\begin{equation}
\hat{H}(f) = (-\text{jsgn}(f)) \hat{Y}(f),    
\end{equation}
where $\text{sgn}(f)$ denotes the sign function and {$j$ denotes the sign of the complex number}. 


The analytic signal (discrete form) is a complex-valued signal that is obtained by the HT:
\begin{equation}
    z(n) = \hat{y}(n) + j\hat{h}(n),
\end{equation}
where $\hat{h}(n) = \mathcal{F}^{-1}(\hat{H}(f))$.


Subsequently, the envelope is defined as the absolute value of $z(n)$, and the ES is the Fourier frequency spectrum of the envelope:
\begin{equation}
    ES(f) = \mathcal{F}(\sqrt{\hat{y}^2(n) + \hat{h}^2(n)}) .
\end{equation}

At last, the objective function is designed to measure the sparsity of the ES as below:
\begin{equation}
\mathcal{L}_f = \frac{\sum_{f=1}^{N}{ES(f)^2}}{(\sum_{f=1}^{N}{ES(f)^4)^{\frac{1}{2}}}}.
\label{eq:lf}
\end{equation}

Notably, this function is similar to the kurtosis, and both these functions are specific versions of the generalized sparsity criterion called $G-l_p/l_q$ norm~\cite{li2014sparsity}:
\begin{equation}
    \mathcal{G}_{l_p/l_q}(x) = sgn\left(\log \left(\frac{q}{p} \right)\right) \frac{\sum_{n=1}^{N}{|x(n)|^p}}{(\sum_{n=1}^{N}{|x(n)|^q)^{\frac{p}{q}}}}, \ p,q>0.
\end{equation}

Interestingly, when $p=4, q=2$, it degeneralizes to the classical kurtosis Eq.~(\eqref{eq:kurt}); when $p=2, q=4$, it becomes Eq.~(~\eqref{eq:lf}). In addition, the values of $p$ and $q$ affect the monotonicity of the objective function (See \textbf{Appendix} for detailed derivation). The derivative of ${G}-{l_p/l_q}$ norm exhibits the following characteristics~\cite{he2021extracting, Liao_2023}:
\begin{equation}
\begin{aligned}
        &\frac{\partial \mathcal{G}_{l_p/l_q}(x)}{\partial x} > 0, \ \mathrm{when} \ p>q>0, \\
        &\frac{\partial \mathcal{G}_{l_p/l_q}(x)}{\partial x} < 0, \ \mathrm{when} \ q>p>0.
\end{aligned}
\end{equation}

This characteristic guides the design of the joint loss function. Comparing Eq.~\eqref{eq:lt} and Eq.~\eqref{eq:lf}, they have opposite signs. As $\mathcal{L}_t$ and $\mathcal{L}_f$ are different in monotonicity, we rearrange them slightly so that they can be optimized towards the same direction. 

\begin{figure}[pos=h]
    \centering
    \includegraphics{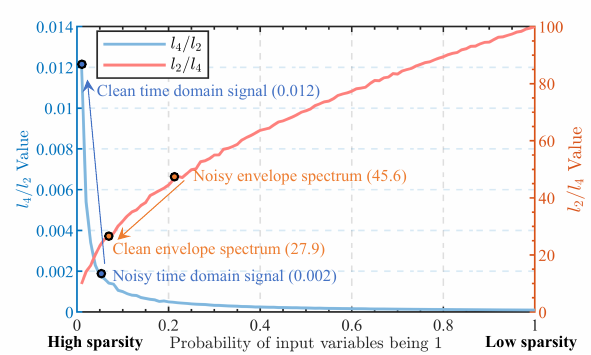}
    \caption{The optimization directions of $l_4/l_2$ and $l_2/l_4$ norm. Where the orange points display the $l_2/l_4$ values of the noisy and clean bearing fault signals in the envelope spectrum. Blue points are the $l_4/l_2$ values of the noisy and clean fault signals in the time domain. The optimization is to maximize the $l_4/l_2$ value of the time domain while minimizing the $l_2/l_4$ value of the frequency domain.}
    \label{fig:glplq}
\end{figure}

We further expound on the differences between the two objective functions, as illustrated in Figure~\ref{fig:glplq}. Initially, we generate Bernoulli distributed random variables $(P(X=1)=p_{0}, \ P(X=0)=1-p_{0}, \ 0<p_{0}<1)$ and feed them into $l_4/l_2$ (kurtosis) and $l_2/l_4$ norms to demonstrate the trends. Although both functions are monotonic, they exhibit opposite directions. More importantly, the $l_4/l_2$ norm follows a more pronounced exponential trend, while the $l_2/l_4$ norm exhibits an approximately linear trend. Secondly, the optimization of two functions in the time domain and ES signals is also predicated on their sparsity. Evidently, the sparsity in the ES is much lower than in the time domain. We anticipate the optimizer to optimize the ES in a linear interval while optimizing the time domain signal in the exponential interval to find their optimal values. Consequently, we establish two distinct functions for BD optimization.

\subsection{Integral optimization with uncertainty-aware weighing scheme}
The fault-diagnosing task typically necessitates a deep learning classifier. Our module exhibits greater flexibility compared to other denoising models as it can be readily transferred to any 1D classifier, such as Transformer and CNN. Upon the addition of the classifier, the loss function evolves into a joint loss:
\begin{equation}
    \mathcal{L} = \mathcal{L}_c + \mathcal{L}_f + \mathcal{L}_t,
\end{equation}
where $\mathcal{L}_t$ and $\mathcal{L}_f$ are derived from Eq.~\eqref{eq:lt} and Eq.~\eqref{eq:lf} respectively; $\mathcal{L}_c$ represents the cross-entropy loss:
\begin{equation}
    \mathcal{L}_c = -\sum_i p_i \log q_i,
\end{equation}
with $p, q$ being the true label and predicted label, respectively.

Our framework seamlessly integrates the objective functions of BD and downstream classifier, thereby incorporating the classification labels as prior information into the BD optimization. Compared to other forms of prior knowledge, such as cyclic frequency~\cite{buzzoni2018blind}, classification labels are more readily obtainable without additional estimation, making them more suitable to guide BD in benefiting the downstream tasks.

In this context, optimizing ClassBD is framed as a multi-task learning problem~\cite{ruder2017overview}. In the context of multi-task learning, a key challenge lies in balancing different loss components. To address this problem, we employ the so-called uncertainty-aware weighing loss to automatically balance the importance of each loss component to the learning problem~\cite{kendall2018multi}. Assume that all the tasks have task-dependent or homoscedastic uncertainty, the loss functions for all tasks are subject to Gaussian noise, then the likelihood function can be defined as:
\begin{equation}
    p(\boldsymbol{y}|f^{\boldsymbol{W}}(\boldsymbol{x})) = \mathcal{N}(f^{\boldsymbol{W}}(\boldsymbol{x}), \sigma^2),
\end{equation}
where $\sigma^2$ denotes the variance of the noise. 

Consequently, the joint loss is formulated as:
\begin{equation}
    \mathcal{L} = \frac{1}{\sigma^2_c}\mathcal{L}_c + \frac{1}{\sigma^2_t}\mathcal{L}_t + \frac{1}{\sigma^2_f}\mathcal{L}_f + \log\sigma_c + \log\sigma_t + \log\sigma_f.
\end{equation}

A large value of $\sigma^2$ will decrease the contribution of the corresponding loss component, and vice versa. Each $\sigma$ is treated as a learnable parameter with value initialized  at $-0.5$ as per~\cite{kendall2018multi, lin2023libmtl}. During training, the scale is regulated by the last term $\log \sigma$, which will be penalized if the $\sigma$ is too large. Although this strategy does not achieve perfect equilibrium, it allows each loss to decrease smoothly and prevents rapid convergence to zero.

\section{Computational experiments}
\subsection{Experimental configurations}
\subsubsection{Signal preprocessing}
In this experiment, we inject additive Gaussian white noise (AWGN) to simulate scenarios with significant noise and validate the classification performance of our method. The level of noise is determined based on the Signal-to-Noise Ratio (SNR), which is defined as follows:
\begin{equation}
\mathrm{SNR} = 10\lg \left(\frac{P_s}{P_n}\right)=20 \lg  \left(\frac{A_s}{A_n}\right),
\end{equation}
where $A_s$ and $A_n$ denote the average amplitude of signal and noise, respectively. An SNR of 0 indicates that the amplitude of the signals is equivalent to that of the noise. The lower the SNR, the higher the noise. 

Under this configuration, we partition the datasets as per~\cite{zhao2020deep}. Firstly, the raw signals are segmented following the time sequence to separate training and test sets, thereby preventing information leakage. Secondly, the sub-sequence is split with or without overlapping sampling, contingent on the volume of the data. Thirdly, we add noise to the datasets at varying SNR levels, which are determined by the performance of models under severe degradation. The noisy signals are then normalized using Z-score standardization. 

It is noteworthy that in our settings, we simulate a more challenging scenario where the noise is generated according to each sub-sequence. In other words, the noise power of each sub-sequence varies. This operation further increases the difficulty of discriminating between similar signals compared to calculating the noise using the entire signal. Finally, for the chosen datasets, the segmentation ratio differs slightly and will be elaborated upon subsequently.


\subsubsection{Baselines and training settings}
We adopt some state-of-the-art time domain bearing fault diagnosis methods as the baselines: 1) Deep residual shrinkage neural network for bearing fault diagnosis (DRSN)~\cite{8850096}; 2) A wavelet convolutional neural network using Laplace wavelet kernel  (WaveletKernelNet)~\cite{9328876}; 3) An enhanced semi-shrinkage wavelet weight initialization network (EWSNet)~\cite{he2023physics}; 4) A Gramian time frequency enhancement network (GTFENet)~\cite{jia2023gtfe}; 5) A time-frequency transform-based neural network (TFN)~\cite{chen2024tfn}. Then, for ClassBD, we adopt WDCNN~\cite{zhang2017new} as our classifier. 

Furthermore, the hyperparameters of all the methods have an identical configuration. All the methods have a maximum training epochs of 200, batch size of 128, learning rate within the range [0.1, 0.3, 0.5, 0.8]. It is noteworthy that we employ SGD~\cite{ruder2016overview} as the optimizer and utilize CosineAnnealingLR~\cite{loshchilov2016sgdr} to dynamically adjust the learning rate throughout the training process. The experiments are executed on a Nvidia RTX 4090 24GB GPU and implemented using Python 3.8 with PyTorch 2.1. All reported results represent the average of ten independent runs.

\subsubsection{Evaluation metrics}
We adopt the commonly used false positive rate (FPR) and F1 score to benchmark the performance of all the considered methods. Formally, these two metrics are defined as below:
\begin{equation}
\nonumber
\begin{aligned}
{\rm FPR} &= \frac{FP}{FP + TN},\\
{\rm Recall} &= \frac{TP}{TP + FN}, \\
{\rm Precision} &= \frac{TP}{TP + FP}, \\
{\rm F1 \ score} &= \frac{\rm 2 \cdot  Precision \cdot Recall}{\rm Precision + Recall},
\end{aligned}
\end{equation}
where TP, TN, FP, and FN stands for the number of true positive, true negative, false positive, and false negative, respectively. 

\subsection{Case study 1: PU dataset}
\subsubsection{Dataset description}

The PU dataset was collected by Paderborn University (PU) Bearing Data Center~\cite{lessmeier2016condition}. This dataset encompasses multiple faults for 32 bearings, including vibration and current signals. The bearings were categorized into three groups: i) Six healthy bearings; ii) Twelve bearings with manually-induced damage (seven with outer race faults, five with inner race faults); iii) Fourteen bearings with real damage, induced by accelerated lifetime tests (five with outer race faults, six with inner race faults, and three with multiple faults). Four operating conditions were implemented, varying the rotational speed (N = 1500rpm or N = 900rpm), load torque (M = 0.7Nm or M = 0.1Nm), and radial force (F = 1000N or F = 400N). The sampling frequency was set at 64KHz. This dataset is one of the more difficult datasets in the field of bearing fault diagnosis~\cite{zhao2020deep}.

In this study, we exclusively utilize the real damaged bearings for classification to validate the methods in real-world scenarios. We classify the fourteen faulty bearings and six healthy bearings into fourteen categories, as illustrated in Table~\ref{tab:puclass}. All the four operating conditions are tested in the experiments, with the codes assigned as N09M07F10, N15M01F10, N15M07F4, and N15M07F10.

Given the large volume of data in the PU dataset, each bearing has 20 segments, so we do not use overlapping to construct the datasets. We slice $20 \times 2048$ sub-segments with a stride of 2048, resulting in 400 data per bearing. The total number of datasets is $7600 \times 2048$. The data collected in the first 19 segments are randomly divided into training and validation sets at a ratio of 0.8:0.2, while the data from the 20th segment are allocated to the test set. Consequently, the dataset includes 5776 training data, 1444 validation data, and 380 test data. Lastly, we established four SNR levels (-4dB, -2dB, 0dB, 2dB) to evaluate the diagnosis performance.

\begin{table}[pos=h]
\centering
\caption{Fourteen categories of PU datasets in our experiments. Where OR and IR denote outer race fault and inner race fault respectively; S, R, and M are single, repetitive, and multiple damages respectively; L1-L3 represent damage levels; F and P are fatigue and plastic deform, respectively.}
\begin{tabular}{@{}lcc|lcc@{}}
\toprule
Category & Bearing Code & Damage         & Category & Bearing Code & Damage         \\ \midrule
1        & K01-K06      & Healthy        & 8        & KB24         & (OR+IR)+M+L3+F \\
2        & KA04         & OR+S+L1+F      & 9        & KB27         & (OR+IR)+M+L1+P \\
3        & KA15         & OR+S+L1+P      & 10       & KI04,KI14    & IR+M+L1+F      \\
4        & KA16         & OR+R+L2+F      & 11       & KI16         & IR+S+L3+F      \\
5        & KA22         & OR+S+L1+F      & 12       & KI17         & IR+R+L1+F      \\
6        & KA30         & OR+R+L1+P      & 13       & KI18         & IR+S+L2+F      \\
7        & KB23         & (OR+IR)+M+L2+F & 14       & KI21         & IR+S+L1+F      \\ \bottomrule
\end{tabular}
\label{tab:puclass}
\end{table}

\subsubsection{Classification results}

The classification results are presented in Table~\ref{tab:puresults}. We highlight key findings from the analysis. Notably, the ClassBD model demonstrates superior performance compared to its competitors across various noise levels and operating conditions. Overall, the average F1 scores of ClassBD on the four conditions are higher than 94\%. Specifically, under high noise conditions (at -4 dB), the ClassBD exhibits a substantial performance gap relative to other methods. For instance, on the N15M01F10 dataset with -4dB noise, the ClassBD achieves an impressive 95\% F1 score, while the second-best method, EWSNet, only attains 70\% F1. These results underscore the efficacy of ClassBD as a robust anti-noise model, consistently delivering competitive performance across diverse high-noise scenarios.

\begin{table}[pos=h]
\centering
\caption{Classification results on the PU datasets. Where bold-faced numbers denote the better results.}
\scalebox{0.85}{
\begin{tabular}{llcccccccc|cc}
\hline
\multirow{2}{*}{Operating Condition} & \multirow{2}{*}{Method} & \multicolumn{2}{c}{SNR=-4dB}                                                 & \multicolumn{2}{c}{SNR=-2dB}         & \multicolumn{2}{c}{SNR=0dB}          & \multicolumn{2}{c|}{SNR=2dB}         & \multicolumn{2}{c}{Average}                                                \\ \cline{3-12} 
                                     &                         & F1 $\uparrow$                                 & FPR $\downarrow$                                & F1 $\uparrow$              & FPR $\downarrow$            & F1  $\uparrow$             & FPR $\downarrow$            & F1 $\uparrow$              & FPR $\downarrow$            & F1 $\uparrow$                                  & FPR $\downarrow$                                \\ \hline
\multirow{6}{*}{N09M07F10}           & WaveletKernelNet        & 67.45\%                              & 2.30\%                              & 74.09\%          & 1.80\%          & 83.14\%          & 1.01\%          & 89.04\%          & 0.68\%          & 78.43\%                              & 1.45\%                              \\
                                     & EWSNet                  & 87.28\%                              & 0.91\%                              & 92.81\%          & 0.49\%          & 96.18\%          & 0.28\%          & 97.60\%          & 0.18\%          & 93.47\%                              & 0.47\%                              \\
                                     & GTFENet                 & 70.68\%                              & 1.80\%                              & 82.19\%          & 1.08\%          & 92.39\%          & 0.52\%          & 90.34\%          & 0.59\%          & 83.90\%                              & 1.00\%                              \\
                                     & TFN                     & 24.24\%                              & 5.87\%                              & 30.12\%          & 5.45\%          & 46.66\%          & 4.37\%          & 3.70\%           & 7.69\%          & 26.18\%                              & 5.85\%                              \\
                                     & DRSN                    & 51.95\%                              & 2.83\%                              & 61.74\%          & 2.28\%          & 71.28\%          & 1.84\%          & 71.42\%          & 1.49\%          & 64.10\%                              & 2.11\%                              \\
                                     & ClassBD                 & \multicolumn{1}{l}{\textbf{92.08\%}} & \multicolumn{1}{l}{\textbf{0.60\%}} & \textbf{95.35\%} & \textbf{0.34\%} & \textbf{98.28\%} & \textbf{0.13\%} & \textbf{99.27\%} & \textbf{0.05\%} & \multicolumn{1}{l}{\textbf{96.25\%}} & \multicolumn{1}{l}{\textbf{0.28\%}} \\ \hline
\multirow{6}{*}{N15M01F10}           & WaveletKernelNet        & 29.91\%                              & 3.94\%                              & 53.56\%          & 2.67\%          & 66.60\%          & 1.95\%          & 82.73\%          & 1.08\%          & 58.20\%                              & 2.41\%                              \\
                                     & EWSNet                  & 69.66\%                              & 1.86\%                              & 82.65\%          & 1.07\%          & 94.89\%          & 0.31\%          & 99.06\%          & 0.06\%          & 86.56\%                              & 0.83\%                              \\
                                     & GTFENet                 & 47.70\%                              & 3.22\%                              & 72.85\%          & 1.51\%          & 92.13\%          & 0.45\%          & 92.89\%          & 0.44\%          & 76.39\%                              & 1.41\%                              \\
                                     & TFN                     & 32.46\%                              & 5.34\%                              & 37.94\%          & 5.01\%          & 51.75\%          & 3.79\%          & 60.85\%          & 3.38\%          & 45.75\%                              & 4.38\%                              \\
                                     & DRSN                    & 52.29\%                              & 2.76\%                              & 67.15\%          & 1.87\%          & 81.88\%          & 1.20\%          & 88.81\%          & 0.78\%          & 72.53\%                              & 1.65\%                              \\
                                     & ClassBD                 & \textbf{95.19\%}                     & \textbf{0.36\%}                     & \textbf{98.87\%} & \textbf{0.08\%} & \textbf{99.63\%} & \textbf{0.02\%} & \textbf{99.71\%} & \textbf{0.02\%} & \textbf{98.35\%}                     & \textbf{0.12\%}                     \\ \hline
\multirow{6}{*}{N15M07F04}            & WaveletKernelNet        & 17.16\%                              & 4.82\%                              & 42.61\%          & 3.30\%          & 75.14\%          & 1.38\%          & 53.78\%          & 2.38\%          & 47.17\%                              & 2.97\%                              \\
                                     & EWSNet                  & 81.06\%                              & 1.12\%                              & 88.49\%          & 0.65\%          & 96.42\%          & 0.23\%          & 98.71\%          & 0.08\%          & 91.17\%                              & 0.52\%                              \\
                                     & GTFENet                 & 65.56\%                              & 1.96\%                              & 78.51\%          & 1.30\%          & 95.60\%          & 0.26\%          & 98.40\%          & 0.10\%          & 84.51\%                              & 0.90\%                              \\
                                     & TFN                     & 26.22\%                              & 5.75\%                              & 31.63\%          & 5.41\%          & 50.40\%          & 3.89\%          & 61.71\%          & 3.24\%          & 42.49\%                              & 4.57\%                              \\
                                     & DRSN                    & 25.49\%                              & 4.14\%                              & 52.14\%          & 2.81\%          & 72.41\%          & 1.56\%          & 87.31\%          & 0.75\%          & 59.34\%                              & 2.32\%                              \\
                                     & ClassBD                 & \textbf{96.52\%}                     & \textbf{0.24\%}                     & \textbf{97.35\%} & \textbf{0.18\%} & \textbf{99.20\%} & \textbf{0.05\%} & \textbf{99.44\%} & \textbf{0.03\%} & \textbf{98.13\%}                     & \textbf{0.13\%}                     \\ \hline
\multirow{6}{*}{N15M07F10}           & WaveletKernelNet        & 45.93\%                              & 3.13\%                              & 51.33\%          & 2.81\%          & 61.39\%          & 2.14\%          & 78.33\%          & 1.18\%          & 59.24\%                              & 2.31\%                              \\
                                     & EWSNet                  & 57.38\%                              & 2.39\%                              & 76.01\%          & 1.37\%          & 95.84\%          & 0.27\%          & 98.60\%          & 0.10\%          & 81.96\%                              & 1.03\%                              \\
                                     & GTFENet                 & 23.93\%                              & 4.26\%                              & 79.42\%          & 1.14\%          & 96.57\%          & 0.21\%          & 97.97\%          & 0.12\%          & 74.47\%                              & 1.43\%                              \\
                                     & TFN                     & 30.47\%                              & 5.44\%                              & 33.87\%          & 5.29\%          & 68.23\%          & 2.88\%          & 8.50\%           & 6.27\%          & 35.27\%                              & 4.97\%                              \\
                                     & DRSN                    & 33.80\%                              & 3.91\%                              & 62.51\%          & 2.23\%          & 86.19\%          & 0.90\%          & 92.14\%          & 0.48\%          & 68.66\%                              & 1.88\%                              \\
                                     & ClassBD                 & \textbf{80.13\%}                     & \textbf{1.28\%}                     & \textbf{97.35\%} & \textbf{0.18\%} & \textbf{99.48\%} & \textbf{0.03\%} & \textbf{99.44\%} & \textbf{0.04\%} & \textbf{94.10\%}                     & \textbf{0.38\%}                     \\ \hline
\end{tabular}}
\label{tab:puresults}
\end{table}

\subsubsection{Classification under small sample conditions}  

In this study, we rigorously assess the performance of various methods under the challenging scenario of extremely limited sample availability. While our approach is not explicitly tailored for small sample issues, we conduct straightforward tests to gain insights into their behavior. 

Specifically, we employ the N09M07F10 dataset, comprising 20 signal segments per category. In the most stringent case, we retain only one sample from each signal segment, reserving the last segment of the signal as the test set. Consequently, we collect one to three samples from each signal segment and finally compose 15, 30, and 45 samples per class for training sets. Notably, we exclude noise from this experiment.

\begin{figure}[pos=h]
    \centering
    \includegraphics{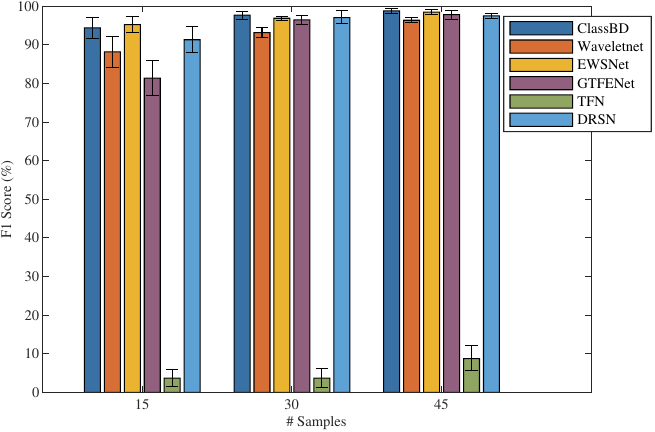}
    \caption{The F1 scores (\%) of baseline methods on the PU N09M07F10 dataset under small sample conditions.} 
    \label{fig:pusmall}
\end{figure}

The results, depicted in Figure~\ref{fig:pusmall}, reveal that even under these constrained conditions, the ClassBD, EWSNet, and DRSN models exhibit commendable performance. Remarkably, all three methods achieve over 90\% F1 scores with a training dataset of just 15 samples per class. Notably, the performance of ClassBD and EWSNet closely align. Their average F1 scores stand at 97.70\% and 97.47\%, respectively, positioning them as the top-performing approaches. In summary, our findings underscore the promising potential of ClassBD in addressing the challenges posed by small sample sizes

\subsection{Case study 2: JNU dataset}
\subsubsection{Dataset description}
This dataset was provided by Jiangnan University (JNU)~\cite{li2013sequential}. Two types of roller bearings were artificially injected with inner race defects (bearing N205), outer race defects and ball defects (bearing NU205) by a wire-cutting machine. Three different rotation speeds (600rpm, 800rpm, 1000rpm) were implemented to test the bearings. The sampling rate was set to 50KHz, resulting in signal segments of 20 seconds each. This dataset is one of the more difficult datasets in the field of bearing fault diagnosis~\cite{zhao2020deep}.

To facilitate varying rotation speed classification, we organized the data into ten classes. These classes encompass three fault types across the three different speeds, along with a healthy class. Furthermore, compared to 
the PU dataset, the JNU dataset is characterized by a limited data volume. Consequently, we adopted an overlapping strategy to extract signal segments, with a stride of 100. The final dataset configuration involved allocating the last 25\% of the data as the test set. Prior to this, the remaining data were randomly partitioned into training and validation sets, maintaining an 80:20 split ratio.

In summary, the dataset statistics are as follows: 54,064 samples in the training set, 13,517 samples in the validation set, and 22,269 samples in the test set. Also, we introduce four SNR levels: -10 dB, -8 dB, -6 dB, and -4 dB for evaluation.

\subsubsection{Classification results}

The classification results, as summarized in Table~\ref{tab:jnuresult}, yield valuable insights. Notably, the ClassBD model consistently outperforms its competitors across various noisy conditions. Overall, The ClassBD achieves an impressive F1 score exceeding 96\%, surpassing the second-best method, EWSNet, which attains an average F1 score of 92.75\%. To be specific, even in the presence of severe noise (at -10 dB), our method maintains a commendable 93\% F1 score. This remarkable performance underscores the ClassBD’s excellent anti-noise capabilities. In comparison, under the same noisy conditions, other methods, excluding EWSNet, experience significant degradation. At last, the JNU dataset experiment substantiates that our approach effectively realizes bearing fault diagnosis across varying rotational speeds, even in challenging high-noise environments.

\begin{table}[pos=h]
\centering
\caption{Classification results on the JNU dataset. Where bold-faced numbers denote the better results.}
\begin{tabular}{@{}lcccccccc|cc@{}}
\toprule
\multirow{2}{*}{Method} & \multicolumn{2}{c}{SNR=-10dB}        & \multicolumn{2}{c}{SNR=-8dB}         & \multicolumn{2}{c}{SNR=-6dB}         & \multicolumn{2}{c|}{SNR=-4dB}        & \multicolumn{2}{c}{Average}        \\ \cmidrule(l){2-11} 
                        & F1 $\uparrow$              & FPR $\downarrow$            & F1  $\uparrow$              & FPR $\downarrow$            & F1 $\uparrow$               & FPR $\downarrow$            & F1 $\uparrow$               & FPR  $\downarrow$           & F1 $\uparrow$               & FPR  $\downarrow$           \\ \midrule
WaveletKernelNet        & 42.19\%          & 4.43\%          & 74.72\%          & 1.66\%          & 70.75\%          & 1.66\%          & 74.36\%          & 1.52\%          & 65.50\%          & 2.32\%          \\
EWSNet                  & 89.39\%          & 0.63\%          & 90.06\%          & 0.62\%          & 95.03\%          & 0.29\%          & 96.51\%          & 0.21\%          & 92.75\%          & 0.44\%          \\
GTFENet                 & 63.35\%          & 2.13\%          & 80.19\%          & 1.28\%          & 91.45\%          & 0.52\%          & 93.98\%          & 0.34\%          & 82.24\%          & 1.07\%          \\
TFN                     & 58.90\%          & 3.01\%          & 39.86\%          & 5.30\%          & 6.69\%           & 10.00\%         & 18.74\%          & 6.58\%          & 31.05\%          & 6.22\%          \\
DRSN                    & 33.60\%          & 4.80\%          & 50.55\%          & 3.36\%          & 56.89\%          & 2.54\%          & 57.96\%          & 2.42\%          & 49.75\%          & 3.28\%          \\
ClassBD                 & \textbf{93.06\%} & \textbf{0.42\%} & \textbf{96.20\%} & \textbf{0.23\%} & \textbf{97.33\%} & \textbf{0.17\%} & \textbf{98.54\%} & \textbf{0.09\%} & \textbf{96.28\%} & \textbf{0.23\%} \\ \bottomrule
\end{tabular}
\label{tab:jnuresult}
\end{table}

\subsection{Case study 3: HIT dataset}
\subsubsection{Dataset description}
\begin{figure}[pos=h]
    \centering
\includegraphics[width=0.8\linewidth]{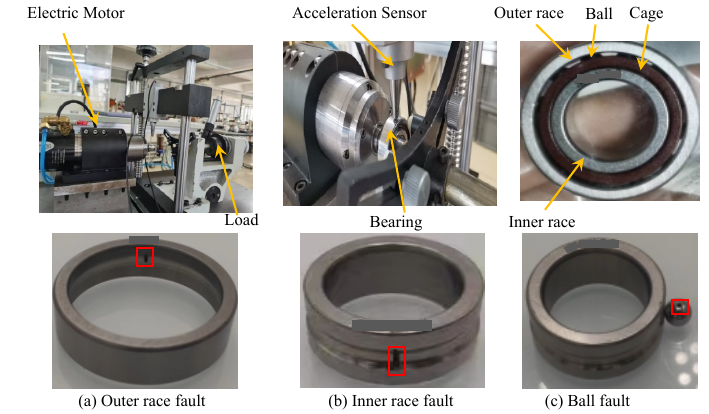}
    \caption{The bearing fault test rig and three types of faults.}
    \label{fig:HITdataset}
\end{figure}

The Harbin Institute of Technology (HIT) dataset was collected by our team. The data acquisition for faulty bearings was conducted at the MIIT Key Laboratory of Aerospace Bearing Technology and Equipment at HIT. The bearing test rig and faulty bearings are illustrated in Figure~\ref{fig:HITdataset}. This test rig follows the Chinese standard rolling bearing measuring method (GB/T 32333-2015). The test bearings used were HC7003 angular contact ball bearings, typically employed in high-speed rotating machines. For signal collection, an acceleration sensor was directly attached to the bearing, and the NI USB-6002 device was used at a sampling rate of 12 KHz. Approximately 47 seconds of data were recorded for each bearing, resulting in 561,152 data points per class. The rotation speed was set to 1800rpm, significantly faster than previous datasets.

Moreover, defects were manually injected at the outer race (OR), inner race (IR), and ball, similar to the JNU dataset. However, we established three severity levels (minor, moderate, severe), resulting in ten classes, as detailed in Table~\ref{tab:class}. This dataset presents a greater challenge than the JNU dataset due to the faults being cracks of the same size but varying depths, leading to more similarity in the features of different categories.

The segmentation of the dataset is similar to that of the JNU dataset. A stride of 28 was set to acquire a larger volume of data. The last quarter of the data was designated as the test set, while the remaining data were randomly divided into training and validation sets at a ratio of 0.8:0.2. Consequently, the training set, validation set, and test set comprise 113920, 28480, and 46732 samples, respectively. Lastly, the SNR levels were set at -10dB, -8dB, -6dB, -4dB.

\begin{table}[pos=h]
\caption{Ten healthy statuses in our HIT dataset. OR and IR denote that the faults appear in the outer race and inner race, respectively.}
\centering
\begin{tabular}{lc|lc}
\toprule 
Category & Faulty Mode        & Category & Faulty Mode  \\ \midrule 
1    & Health                                    & 6    & OR (Moderate)  \\ 
2    & Ball cracking (Minor)       & 7    & OR (Severe)   \\ 
3    & Ball cracking (Moderate)     & 8    & IR (Minor)   \\ 
4    & Ball cracking (Severe)       & 9    & IR (Moderate) \\ 
5    & OR cracking (Minor) & 10   & IR (Severe)   \\ \bottomrule
\end{tabular}
\label{tab:class}
\end{table}

\subsubsection{Classification results}
The classification results are presented in Table~\ref{tab:hitresult}. Several observations observations are as follows. Firstly, ClassBD outperforms its competitors in terms of both F1 score and FPR across all noise conditions. The F1 score of ClassBD exceeds 96\%, compared to the second-best method, EWSNet, which achieves an average F1 score of 92.75\%. Secondly, it is noteworthy that even under severe noise conditions (-10dB), ClassBD maintains an F1 score of 93\%, indicating its robust anti-noise performance. Under these conditions, the performance of all other methods, except EWSNet, degrades significantly. This phenomenon highlights the superiority of ClassBD.

\begin{table}[pos=h]
\centering
\caption{Classification results on the HIT dataset. Where bold-faced numbers denote the better results.}
\begin{tabular}{@{}lcccccccc|cc@{}}
\toprule
\multirow{2}{*}{Method} & \multicolumn{2}{c}{SNR=-10dB}        & \multicolumn{2}{c}{SNR=-8dB}         & \multicolumn{2}{c}{SNR=-6dB}         & \multicolumn{2}{c|}{SNR=-4dB}        & \multicolumn{2}{c}{Average}        \\ \cmidrule(l){2-11} 
                        & F1 $\uparrow$               & FPR $\downarrow$             & F1 $\uparrow$               & FPR $\downarrow$            & F1 $\uparrow$               & FPR $\downarrow$            & F1 $\uparrow$               & FPR $\downarrow$            & F1  $\uparrow$              & FPR $\downarrow$            \\ \midrule
WaveletKernelNet        & 20.93\%          & 7.68\%          & 51.64\%          & 4.64\%          & 66.58\%          & 3.17\%          & 83.89\%          & 1.49\%          & 55.76\%          & 4.25\%          \\
EWSNet                  & 48.23\%          & 5.10\%          & 63.91\%          & 3.58\%          & 78.39\%          & 2.02\%          & 83.99\%          & 1.49\%          & 68.63\%          & 3.05\%          \\
GTFENet                 & 42.01\%          & 5.63\%          & 52.53\%          & 4.93\%          & 76.79\%          & 2.58\%          & 82.90\%          & 1.68\%          & 63.56\%          & 3.70\%          \\
TFN                     & 7.52\%           & 9.53\%          & 19.90\%          & 8.54\%          & 36.14\%          & 6.67\%          & 61.65\%          & 4.02\%          & 31.30\%          & 7.19\%          \\
DRSN                    & 32.83\%          & 6.80\%          & 61.67\%          & 3.77\%          & 59.31\%          & 4.12\%          & 84.85\%          & 1.38\%          & 59.67\%          & 4.01\%          \\
ClassBD                 & \textbf{86.15\%} & \textbf{1.29\%} & \textbf{86.84\%} & \textbf{1.42\%} & \textbf{93.74\%} & \textbf{0.69\%} & \textbf{97.23\%} & \textbf{0.32\%} & \textbf{90.99\%} & \textbf{0.93\%} \\ \bottomrule
\end{tabular}
\label{tab:hitresult}
\end{table}

\begin{figure}[pos=h]
    \centering
    \includegraphics[width=\linewidth]{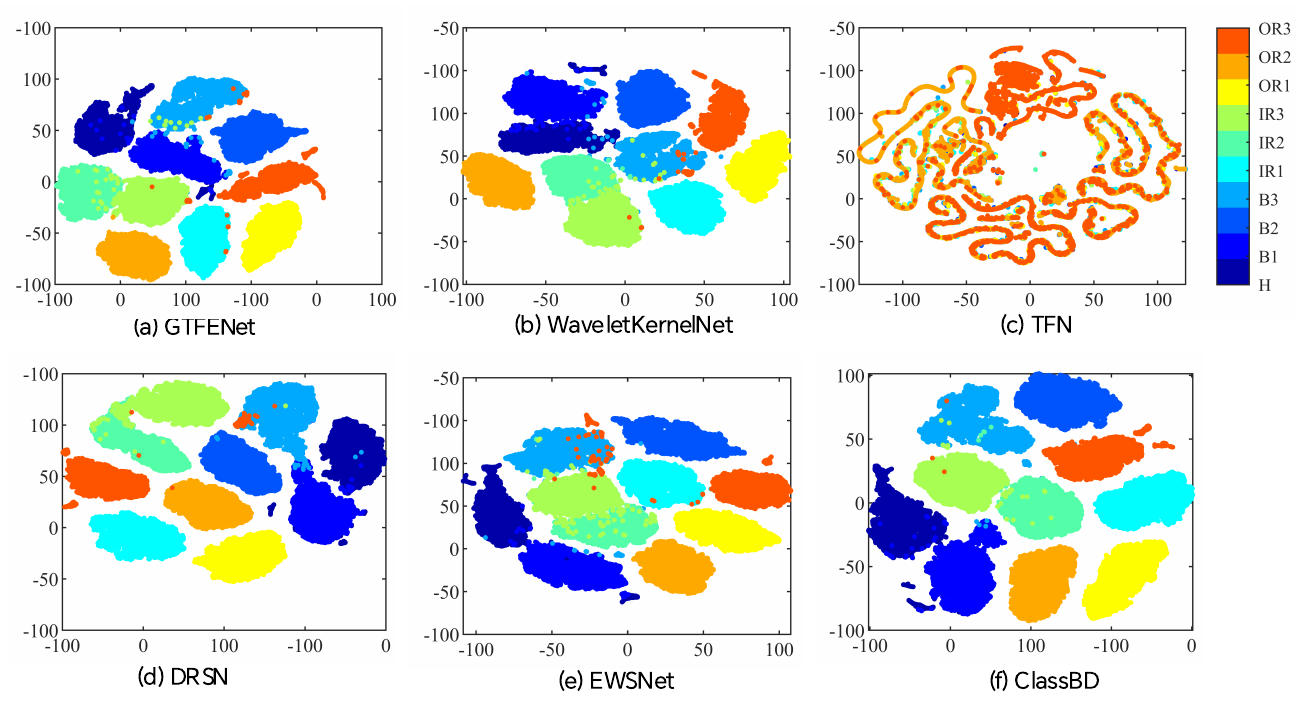}
    \caption{The t-SNE visualization of the last convolutional features of all methods on the -10dB HIT dataset.}
    \label{fig:tsne}
\end{figure}

Furthermore, we reduce the dimensions of the last layer features of all methods to 2D space using t-SNE~\cite{tsne} under -10dB noise. The results are depicted in Figure~\ref{fig:tsne}. Overall, the features of the five methods can generate distinct clusters, with the exception of TFN, which aligns with their categorical performance. However, all methods exhibit some degree of misclassification. For instance, the red points (OR3) of DRSN and EWSNet appear within the blue clusters (B3), indicating that some instances of OR3 are misclassified as B3. Similarly, in GTFENet and WaveleKernelNet, some green points (IR3) overlap with the IR2 and B3 clusters. In contrast, ClassBD only has a few outliers that are misclassified into other clusters, demonstrating its superior feature extraction ability under high noise conditions.

\begin{figure}[pos=h]
    \centering
    \includegraphics[scale=0.9]{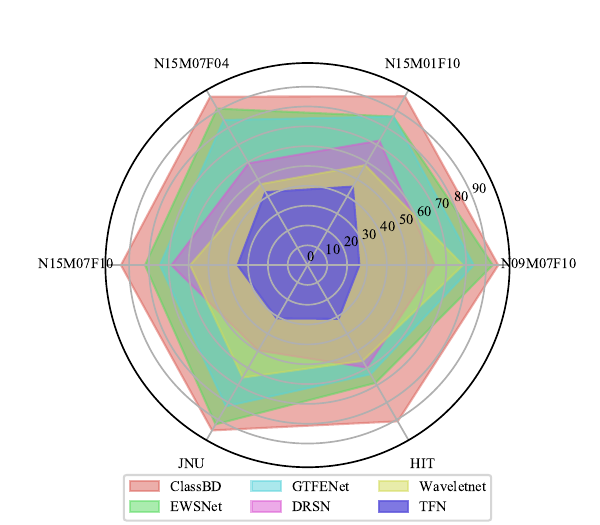}
    \caption{The average F1 scores of all methods across the compared datasets.}
    \label{fig:radar}
\end{figure}

In summary, based on the performance across three datasets, we illustrate the average F1 scores of all methods in Figure~\ref{fig:radar}. The results indicate that our method, ClassBD, outperforms other baseline methods on all datasets, with average F1 scores exceeding 90\%. This underscores the robust anti-noise performance of ClassBD.

\section{Computational experiments}

\subsection{Comparison of BD methods}

Despite the recent proposal of many BD methods, a few works have validated their fault diagnosis performance via the integration of classifiers. Initially, we assess the performance of these BD methods by employing WDCNN as the post-BD classifier. For comparative analysis, we utilize four prior-free BD methods: minimum entropy deconvolution (MED)~\cite{wiggins1978minimum}, sparse maximum harmonics-to-noise-ratio Deconvolution (SMHD)~\cite{miao2016sparse}, improved maximum correlated kurtosis deconvolution (IMCKD)~\cite{miao2017application}, and multi-task neural network blind deconvolution (MNNBD)~\cite{Liao_2023}. The metric employed for this evaluation is also the F1 score.

The classification results of the three datasets are presented in Table~\ref{tab:CompareBD}. Collectively, our approach outperforms others on all three datasets, indicating that the classifier-guided optimization is more apt for fault diagnosis. 
This is because traditional BD methods employ unsupervised optimization, whereas our method implements supervised learning, which introduces class-aware information. Furthermore, our approach is a streamlined one-step pipeline. When compared to the combination of conventional BD and classifiers, our method demonstrates superior effectiveness under noise-affected conditions.

\begin{table}[pos=h]
\caption{The F1 scores (\%) of different BD methods on three datasets.}
\centering
\begin{tabular}{@{}llllll@{}}
\toprule
      & \multicolumn{1}{c}{MED} & \multicolumn{1}{c}{SHMD} & \multicolumn{1}{c}{IMCKD} & \multicolumn{1}{c}{MNNBD} & \multicolumn{1}{c}{ClassBD} \\ \midrule
N09M07F10 0dB   & 63.21\%                 & 75.12\%                  & 73.35\%                   & 72.52\%                   & \textbf{98.28\%}                     \\
JNU -6dB & 88.64\%                 & 89.88\%                  & 90.02\%                   & 90.33\%                   & \textbf{97.51\%}                     \\
HIT -6dB & 66.12\%                 & 70.61\%                  & 69.63\%                   & 73.23\%                   & \textbf{98.47\%}                     \\ \bottomrule
\end{tabular}
\label{tab:CompareBD}
\end{table}

On the other hand,  it is imperative to validate the denoising performance of BD methods in the frequency domain, thereby demonstrating that these methods indeed extract fault-related features and provide interpretability. Therefore, we design a frequency domain metric called fault frequency index (FFI), which is defined as follows:
\begin{equation}
    {\rm FFI}= \frac{1}{I}\sum_{i=1}^{I}\max [EES(if_c-0.1f_c), EES(if_c+0.1f_c) ].
\end{equation}
Where, $f_c$ represents the fault characteristic frequency, $I$ is the number of harmonic frequencies (set to $I = 5$), $\pm 0.1$ constitutes an interval of the random drift of $f_c$ in the actual measured signals~\cite{antoni2006spectral}, The term $EES(\cdot)$ signifies the enhanced envelope spectrum, obtained by performing Fast-SC~\cite{antoni2017fast}: 
\begin{equation}
    EES(\alpha) = \frac{1}{N}\sum_{f=1}^{N}|\gamma_x(\alpha,f)|,
\end{equation}
where $\gamma_x(\alpha,f)$ is the spectral coherence, $N$ denotes the number of discrete frequency, and $\alpha$ is the cyclic frequency, which encompasses fault characteristic frequencies $f_c$ and other cyclic frequency components. Compared to the squared envelope spectrum (ES), the EES can amplify the non-zero cyclic frequency. Hence, the EES is apt for identifying the amplitude of fault characteristic frequencies and computing the FFI.

Subsequently, we utilize the JNU dataset to demonstrate the feature extraction performance. The SNR is set to -10dB, and we apply BD methods to all faulty signals, and then calculate the FFI of the signals post-BD. The results are illustrated in Table~\ref{tab:bdffi}. From these results, we can make several observations. Firstly, ClassBD is highly effective in extracting fault-related features, as evidenced by the highest FFIs across all signals, surpassing even the raw clean signals. Secondly, on average, all BD methods are capable of enhancing the fault characteristic frequencies. The FFIs are consistently higher than the noisy signals, with MNNBD ranking second-best, achieving an average improvement of 41\% compared to the noisy signals.

\begin{table}[pos=h]
\caption{The FFI of different BD methods on the JNU dataset. Higher is better. OR, IR, B represent the outer, inner and ball faults respectively and the numbers (600,800, 1000) denote the rotating speeds.}
\begin{tabular}{@{}lccccccccc|c@{}}
\toprule
                     & OR600 & OR800 & OR1000 & IR600 & IR800 & IR1000 & B600 & B800 & B1000                     & Average       \\ \midrule
Raw signal           & 0.51  & 0.38  & 0.38   & 0.38  & 0.45  & 0.39   & 0.31 & 0.60 & \multicolumn{1}{c|}{0.49} & 0.43          \\
Noisy signal (-10dB) & 0.25  & 0.17  & 0.16   & 0.17  & 0.24  & 0.20   & 0.11 & 0.10 & \multicolumn{1}{c|}{0.09} & 0.17          \\
IMCKD                & 0.25  & 0.18  & 0.15   & 0.18  & 0.22  & 0.19   & 0.12 & 0.25 & \multicolumn{1}{c|}{0.19} & 0.19          \\
SHMD                 & 0.18  & 0.14  & 0.11   & 0.17  & 0.16  & 0.21   & 0.19 & 0.25 & \multicolumn{1}{c|}{0.23} & 0.18          \\
MED                  & 0.25  & 0.18  & 0.15   & 0.18  & 0.22  & 0.19   & 0.12 & 0.26 & \multicolumn{1}{c|}{0.19} & 0.19          \\
MNNBD                & 0.13  & 0.23  & 0.29   & 0.24  & 0.31  & 0.28   & 0.20 & 0.17 & \multicolumn{1}{c|}{0.29} & 0.24          \\
ClassBD              & 0.62  & 0.42  & 0.44   & 0.68  & 0.86  & 0.73   & 0.37 & 0.87 & 0.48                      & \textbf{0.61} \\ \bottomrule
\end{tabular}
\label{tab:bdffi}
\end{table}

\begin{figure}[pos=h]
    \centering
    \includegraphics[width=\linewidth]{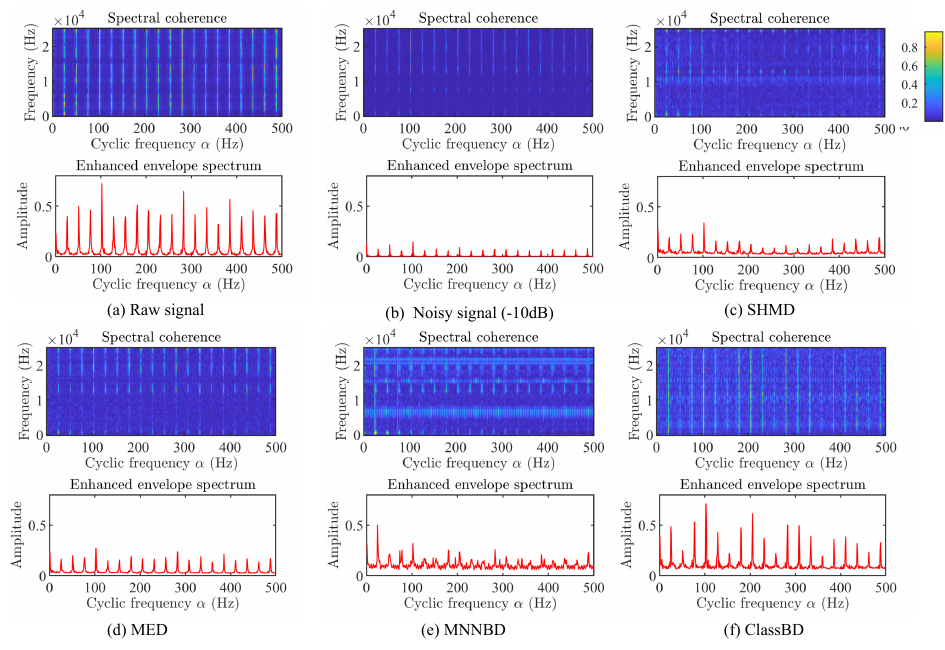}
    \caption{Comparison of BD methods on the JNU B1000 signal.}
    \label{fig:bdcompare}
\end{figure}

Finally, we present a comparison of the feature extraction performance on the B1000 signal using Fast-SC. As depicted in Figure~\ref{fig:bdcompare}, the bright lines in the spectral coherence suggest that the magnitudes of the cyclic frequencies $\alpha$ and corresponding frequency bands are elevated. 
And the EES displays the intensity of the extracted cyclic frequencies of the signal post-BD. Several observations can be made from this. Firstly, the noise significantly attenuates the characteristics of the signal. A comparison between Figure~\ref{fig:bdcompare} (a) and (b) reveals that the intensity of all frequency bands is suppressed, with the lower frequency bands being more severely drown out by the noise. Secondly, all BD methods are capable of enhancing the cyclic frequency characteristics of the signal, indicating the effectiveness of BD in extracting fault-related features from noisy signals. Specifically, MNNBD focuses on the fundamental cyclic frequency, resulting in the extraction of full frequency bands at the first cyclic frequency. While SHMD and MED enhance all cyclic frequency bands, the amplitude of the individual cyclic frequencies is lower. Lastly, ClassBD exhibits the best performance. It significantly enhances the cyclic frequency characteristics of the signal. More importantly, ClassBD can provide valuable interpretability in decision-making.

\subsection{Classification results on various noise conditions}
To underscore the superior performance of ClassBD in handling various types of noise, we conduct a comparative analysis using several synthetic and real-world noises.

First, we generate Gaussian and Laplace noises from their respective probability density functions:
\begin{equation}
\begin{aligned}
        {\rm Gaussian \ Noise:} \ &g(x|\mu, \sigma) = \frac{1}{\sigma\sqrt{2\pi}}e^{\frac{-(x-\mu)^2}{2\sigma^2}}, \\
        {\rm Laplace \ Noise:} \ &l(x|\gamma, b) = \frac{1}{2b}e^{\frac{-|x-\gamma|}{b}},
\end{aligned}
\end{equation}
where $\mu$ and $\sigma$ denote mean and standard deviation respectively, and are set $\mu=0, \sigma=1$ for Gaussian noise. For Laplace noise, $\gamma$ represents the location parameter and $b$ is the scale parameter, with $\gamma=0, b=1$. 

Furthermore, pink noise is generated in the frequency domain, with its power spectral density inversely proportional to the signal frequency:

\begin{equation}
    {\rm Pink \ Noise:} \ S(f) \propto \frac{1}{f^{\alpha}},
\end{equation}
where $f$ denotes the randomly determined frequency of the noise, and $\alpha$ is set to 1. In this experiment, we validate the classification performance on the PU "N09M07F10" dataset and set the SNR to -4dB for all synthetic noises.

Lastly, we employ two types of real-world noises, collected from an airplane and a truck using acoustic sensors, to simulate bearings operating under real-world conditions~\footnote{\url{https://github.com/markostam/active-noise-cancellation}}. Due to the discrepancy between the signal power captured by the acoustic and vibration sensors, we adjust the noise power to match the power of the vibration signal, thereby simulating a real-world bearing failure scenario.

The experimental results, presented in Table~\ref{tab:comparenoise}, reveal that our method delivers the highest average F1 score, outperforming the second-best method by 20\%. This underscores the superior anti-noise performance of ClassBD across different types of noise. Specifically, Gaussian noise is the easiest to handle, with several methods (WaveletKernelNet, EWSNet, GTFENet, ClassBD) achieving optimal results at the same SNR. Conversely, pink noise appears to be the most challenging, as evidenced by the highest-performing method only achieving an 83\% F1 score. This is likely due to pink noise being generated in the frequency domain, which can interfere with the signal frequency. Finally, when considering the two real-world noises, ClassBD significantly outperforms other methods, demonstrating approximately 40\% improvement over the second-best method. A comparison of the performance of several methods reveals that the difficulty of handling real-world noise lies somewhere between Gaussian and Laplace noise, indicating the efficacy of using synthetic noise to simulate real-world scenarios.

\begin{table}[pos=h]
\centering
\caption{The F1 scores (\%) of all methods on the PU "N09M07F10" dataset with different noise.}
\begin{tabular}{@{}lcccccc@{}}
\toprule
                 & Airplane         & Truck            & Pink=-4dB        & Laplace=-4dB     & Gaussian=-4dB    & Average          \\ \midrule
WaveletKernelNet & 49.88\%          & 47.94\%          & 48.36\%          & 59.08\%          & 67.45\%          & 54.54\%          \\
EWSNet           & 61.19\%          & 58.30\%          & 80.58\% & 60.51\%          & 87.28\%          & 69.57\%          \\
GTFENet          & 61.19\%          & 67.36\%          & 68.58\%          & 41.27\%          & 70.68\%          & 61.81\%          \\
TFN              & 26.64\%          & 26.07\%          & 24.05\%          & 14.39\%          & 24.24\%          & 23.07\%          \\
DRSN             & 58.96\%          & 55.36\%          & 48.13\%          & 51.49\%          & 51.95\%          & 53.17\%          \\
ClassBD          & \textbf{90.01\%} & \textbf{91.01\%} & \textbf{83.68\%}          & \textbf{90.52\%} & \textbf{92.08\%} & \textbf{89.46\%} \\ \bottomrule
\end{tabular}
\label{tab:comparenoise}
\end{table}

\subsection{Employing ClassBD to deep learning classifiers}

Given that ClassBD serves as a signal preprocessing module, it possesses the flexibility to be integrated into various backbone networks. In the context of this experiment, we assess the classification performance by incorporating ClassBD into four widely recognized deep learning classifiers: ResNet~\cite{he2016deep}, MobileNetV3~\cite{Howard_2019_ICCV}, WDCNN~\cite{zhang2017new}, and Transformer~\cite{vaswani2017attention}. It is important to note that some networks were initially designed for image classification, so some parameters are revised to accommodate the 1D signal input. The properties of these backbone networks are illustrated in Table~\ref{tab:backbone}.


\begin{table}[pos=h]
\centering
\caption{The properties of compared backbone networks. Where \#FLOPs represents floating point operations, inference time denotes the elapsed time to infer one sample ($1 \times 2048$) on an Intel i9-10900K CPU.}
\begin{tabular}{@{}lccc@{}}
\toprule
Backbone network & \#Params & \#FLOPs & Inference time (ms)\\ \midrule
WDCNN            & 67.30k   & 1.61M   & 3.99           \\
Transformer      & 0.41M    & 7.45M   & 5.98           \\
ResNet18         & 3.83M    & 0.35G   & 13.96          \\ 
MobilieNetV3     & 5.32M    & 77.4M   & 22.98          \\ \bottomrule
\end{tabular}
\label{tab:backbone}
\end{table}

Firstly, the classification results on the PU dataset are illustrated in Table~\ref{tab:backbonePU}. Predominantly, ClassBD has demonstrated its efficacy in enhancing performance. For instance, in the case of N09M07F10 with -4dB noise, the application of ClassBD results in an improvement exceeding 10\% across all backbones. On average, when the SNR is -4dB, ClassBD achieves F1 scores of 87.60\%, whereas the raw networks only yield 64.07\%.

\begin{table}[pos=h]
\centering
\caption{The F1 scores (\%) of the compared backbone networks on the PU datasets.}
\begin{tabular}{@{}llcccccc@{}}
\toprule
\multirow{2}{*}{Working Condition} & \multirow{2}{*}{Backbone Network} & \multicolumn{3}{c}{SNR=-4dB}      & \multicolumn{3}{c}{SNR=-2dB}      \\ \cmidrule(l){3-8} 
                                   &                                   & Raw     & ClassBD & Improvement & Raw     & ClassBD & Improvement \\ \midrule
\multirow{4}{*}{N09M07F10}         & MobilieNetV3                      & 71.86\% & 84.49\% & 12.64\%     & 77.84\% & 97.20\% & 19.36\%     \\
                                   & ResNET18                          & 63.39\% & 84.99\% & 21.60\%     & 73.11\% & 74.29\% & 1.17\%      \\
                                   & WDCNN                             & 50.75\% & 79.06\% & 28.31\%     & 65.74\% & 95.35\% & 29.61\%     \\
                                   & Transformer                       & 61.15\% & 82.34\% & 21.19\%     & 64.87\% & 87.69\% & 22.82\%     \\ \midrule
\multirow{4}{*}{N15M01F10}         & MobilieNetV3                      & 84.30\% & 82.65\% & -1.66\%     & 91.32\% & 96.15\% & 4.83\%      \\
                                   & ResNET18                          & 3.70\%  & 88.57\% & 84.88\%     & 22.78\% & 94.22\% & 71.44\%     \\
                                   & WDCNN                             & 78.67\% & 95.19\% & 16.51\%     & 90.71\% & 98.87\% & 8.16\%      \\
                                   & Transformer                       & 75.94\% & 81.45\% & 5.51\%      & 82.76\% & 89.29\% & 6.54\%      \\ \midrule
\multirow{4}{*}{N15M07F04}         & MobilieNetV3                      & 86.95\% & 94.55\% & 7.60\%      & 91.37\% & 97.82\% & 6.45\%      \\
                                   & ResNET18                          & 40.57\% & 87.28\% & 46.71\%     & 56.53\% & 96.46\% & 39.93\%     \\
                                   & WDCNN                             & 81.59\% & 96.52\% & 14.94\%     & 92.60\% & 97.35\% & 4.75\%      \\
                                   & Transformer                       & 83.68\% & 88.37\% & 4.69\%      & 79.41\% & 80.34\% & 0.93\%      \\ \midrule
\multirow{4}{*}{N15M07F10}         & MobilieNetV3                      & 89.19\% & 91.99\% & 2.80\%      & 91.80\% & 97.66\% & 5.86\%      \\
                                   & ResNET18                          & 3.68\%  & 94.05\% & 90.37\%     & 88.89\% & 95.41\% & 6.52\%      \\
                                   & WDCNN                             & 74.35\% & 80.13\% & 5.78\%      & 59.33\% & 97.35\% & 38.02\%     \\
                                   & Transformer                       & 75.35\% & 89.97\% & 14.62\%     & 76.97\% & 94.88\% & 17.91\%     \\ \midrule
Average                            &                                   & 64.07\% & 87.60\% & 23.53\%     & 75.38\% & 93.15\% & 17.77\%     \\ \bottomrule
\end{tabular}
\label{tab:backbonePU}
\end{table}
Subsequently, we also test these models to the JNU and HIT datasets, setting the SNRs to -6dB and -4dB respectively. As depicted in Table~\ref{tab:backboneJNU}, ClassBD exhibits commendable performance on both datasets, and all backbones experience a performance boost. Specifically, on the JNU dataset, the F1 scores exceed 90\% post the employment of ClassBD, irrespective of the backbone performance. Furthermore, a substantial improvement is observed in the Transformer. The raw Transformer initially yields F1 scores around 50\% on the JNU and HIT datasets, which, after the application of ClassBD, escalates to an average F1 score of 90\%.

\begin{table}[pos=h]
\centering
\caption{The F1 scores (\%) of the compared backbone networks on the JNU and HIT datasets.}
\begin{tabular}{@{}llcccccc@{}}
\toprule
\multirow{2}{*}{Dataset} & \multirow{2}{*}{Backbone Network} & \multicolumn{3}{c}{SNR=-6dB}      & \multicolumn{3}{c}{SNR=-4dB}      \\ \cmidrule(l){3-8} 
                         &                                   & Raw     & ClassBD & Improvement & Raw     & ClassBD & Improvement \\ \midrule
\multirow{4}{*}{JNU}     & MobilieNetV3                      & 84.16\% & 93.59\% & 9.42\%      & 90.61\% & 96.73\% & 6.12\%      \\
                         & ResNET18                          & 69.73\% & 95.31\% & 25.59\%     & 6.69\%  & 97.73\% & 91.04\%     \\
                         & WDCNN                             & 94.35\% & 97.33\% & 2.98\%      & 95.98\% & 98.54\% & 2.57\%      \\
                         & Transformer                       & 54.25\% & 94.58\% & 40.32\%     & 53.30\% & 96.81\% & 43.51\%     \\ \midrule
Average                  &                                   & 75.62\% & 95.20\% & 19.58\%     & 61.65\% & 97.45\% & 35.81\%     \\ \midrule
\multirow{4}{*}{HIT}     & MobilieNetV3                      & 68.62\% & 94.31\% & 25.69\%     & 86.21\% & 95.38\% & 9.17\%      \\
                         & ResNET18                          & 49.24\% & 86.18\% & 36.94\%     & 85.03\% & 86.57\% & 1.53\%      \\
                         & WDCNN                             & 85.11\% & 93.74\% & 8.63\%      & 94.15\% & 97.23\% & 3.07\%      \\
                         & Transformer                       & 42.95\% & 80.14\% & 18.26\%     & 61.88\% & 95.58\% & 52.63\%     \\ \midrule
Average                  &                                   & 61.48\% & 88.59\% & 22.38\%     & 81.82\% & 93.69\% & 16.60\%     \\ \bottomrule
\end{tabular}
\label{tab:backboneJNU}
\end{table}

Finally, the results on the three datasets suggest that ClassBD can function as a plug-and-play denoising module, thereby enhancing the performance of deep learning classifiers under substantial noise. Considering both performance and model size, the simplest backbone, WDCNN, achieves consistent performance under all conditions. Consequently, we recommend it as the backbone for bearing fault diagnosis.

\subsection{Employing ClassBD to machine learning classifiers}

In comparison of deep learning models, classical machine learning (ML) classifiers offer some distinct advantages, including robust interpretability and lightweight models. However, these “shallow” ML methods invariably rely on human-engineered features for bearing fault diagnosis, thereby exhibiting limited generalization ability in the design of end-to-end diagnozing models~\cite{zhang2020deep}. Given that ClassBD can enhance the performance of deep learning classifiers, we posit that it can also serve as a feature extractor to augment the performance of classical ML classifiers. 

Consequently, in this experiment, we utilize the pre-trained ClassBD as a feature extractor and feed the output of ClassBD into several ML classifiers for comparison: support vector machine (SVM)~\cite{cortes1995support}, k-nearest neighbor (KNN)~\cite{mucherino2009k}, random forest (RF)~\cite{ho1995random}, logistic regression (LR)~\cite{cox1958regression}, and a highly efficient gradient boosting decision tree (LightGBM)~\cite{ke2017lightgbm}.

The results are presented in Table~\ref{tab:ml}. Evidently, ClassBD significantly facilitates the performance of ML methods. Directly inputting raw signals into these ML classifiers results in markedly poor performance. On the JNU and HIT datasets, SVM and RF even fail to converge. However, with the incorporation of ClassBD, the classification performance experiences a substantial improvement. For instance, the KNN achieves a 90.71\% F1 score on the JNU dataset, compared to a mere 2.89\% F1 score without ClassBD. This performance even surpasses some deep learning methods. Nonetheless, ML methods exhibit instability across different datasets. The best-performing ML methods can only achieve a 49.77\% score on the PU dataset. Despite this, we believe that the combination of ClassBD and ML methods presents a promising solution, promoting the study of high interpretability and efficiency in diagnostic approaches. We will explore this topic further in our future work.

\begin{table}[pos=h]
\centering
\caption{The F1 scores (\%) of different machine learning methods on three datasets. Where the bold-faced numbers are the better results, '-' denotes the model failed to converge.}
\scalebox{0.9}{
\begin{tabular}{@{}lcccccccccc@{}}
\toprule
\multirow{2}{*}{Dataset} & \multicolumn{2}{c}{SVM} & \multicolumn{2}{c}{KNN}   & \multicolumn{2}{c}{RF} & \multicolumn{2}{c}{LR} & \multicolumn{2}{c}{LightGBM} \\ \cmidrule(l){2-11} 
                         & Raw        & ClassBD    & Raw    & ClassBD          & Raw        & ClassBD   & Raw        & ClassBD   & Raw      & ClassBD           \\ \midrule
PU-N09M07F10 (-4dB)        & 16.59\%    & 40.28\%    & 6.80\% & \textbf{49.77\%} & 7.70\%     & 31.99\%   & 15.48\%    & 52.73\%   & 12.30\%  & 38.68\%           \\
JNU (-10dB)                & -     & 89.74\%    & 2.89\% & \textbf{90.71\%} & 16.28\%    & 90.30\%   & 13.58\%    & 88.06\%   & 18.36\%  & 89.38\%           \\
HIT (-10dB)                 & -     & 72.58\%    & 1.90\% & 73.69\%          & -     & 68.91\%   & 23.85\%    & 49.28\%   & 35.82\%  & \textbf{75.78\%}  \\ \bottomrule
\end{tabular}}
\label{tab:ml}
\end{table}

\subsection{Feature extraction ability of quadratic and conventional networks}
\label{sec:qvc}
{Previously, we have theoretically demonstrated that quadratic networks possess superior cyclostationary feature extraction ability to conventional networks. It is also necessary to validate the performance in practice. Therefore, we construct two time-domain filters using quadratic convolutional layers and conventional convolutional layers with an identical structure and then evaluate their feature-extraction performance on the JNU dataset subjected to -10 dB noise.}

\begin{figure}[pos=h]
    \centering
    \includegraphics{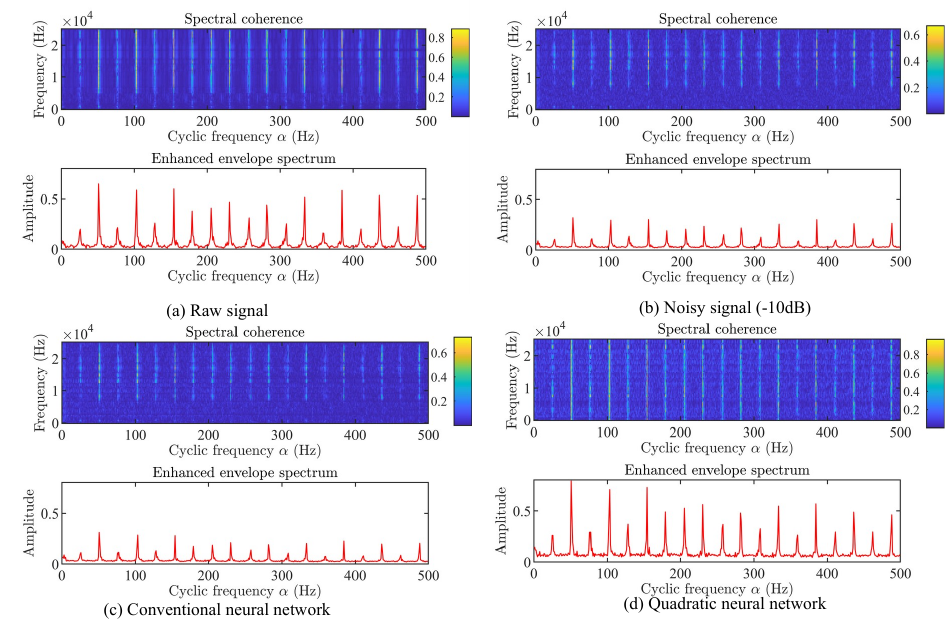}
    \caption{Denosing performance of the quadratic neural network and conventional network on the JNU dataset.}
    \label{fig:qvc}
\end{figure}

{The signals are analyzed using the Fast-SC method~\cite{antoni2017fast}. The results, as depicted in Figure~\ref{fig:qvc}, clearly demonstrate that the quadratic network outperforms in terms of feature extraction capability. The bright lines in the spectral coherence, highlight the quadratic network can extract cyclic frequency across high and low frequency bands. Despite the severe attenuation of the signal amplitude due to the noise, the quadratic network effectively recovers the cyclic frequency of the signal. Remarkably, the amplitude of the initial few cyclic frequencies is even higher than that of the raw signal.}

\subsection{Comparison of ClassBD filters}

Given that ClassBD comprises two filters, we explore their respective contributions in this experiment. Here we test four combinations: a standalone time domain filter (T-filter), a standalone frequency domain filter (F-filter), an F-filter followed by a T-filter, and our proposed scheme (a T-filter followed by an F-filter). 

The results are presented in Table~\ref{tab:comparefilter}. Primarily, our scheme exhibits superior performance, indicating that both filters contribute significantly to the classification. Secondly, when comparing the single T-filter and F-filter, their performances are found to be dataset-dependent. On the PU and HIT datasets, T-filters outperform F-filters, whereas their performance is inferior to F-filters on the JNU dataset.

\begin{figure}[pos=h]
    \centering
    \includegraphics{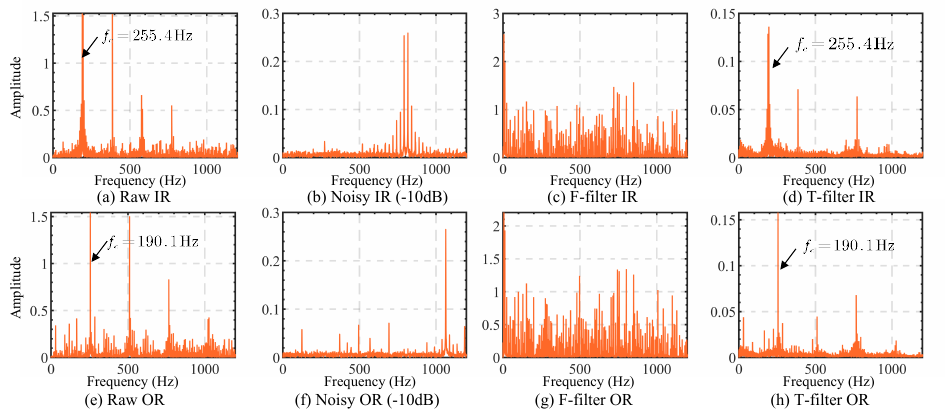}
    \caption{The envelope spectrum of the signals and the output of F/T filters on the HIT dataset. Where IR denotes inner race fault, OR denotes outer race fault, and  $f_c$ is the fault characteristic frequency.}
    \label{fig:filter}
\end{figure}

Lastly, the F-T filter demonstrates instability across the three datasets, with the average F1 score showing only about a 1\% improvement compared to the single filter module. A plausible explanation for this is that the F-filter operates across the entire frequency domain. If it is employed as the first filter, it is susceptible to significant information loss. This is illustrated in Figure~\ref{fig:filter}, where we plot the envelope spectrum of the outputs of the F-filter and T-filter. A comparison of raw and noisy signals reveals that the fault characteristic frequencies $f_c$ are severely suppressed and distorted. While T-filters can still recover the $f_c$ in different faults, the F-filter amplifies the energy across all frequency bands, causing the $f_c$ to be overwhelmed. Therefore, our scheme initially employs a T-filter to extract $f_c$ from the noisy signals, followed by an F-filter to enhance the energy in the frequency domain.

\begin{table}[pos=h]
\caption{The F1 scores (\%) of different neural filters on three datasets. Where T-filter represents time domain quadratic convolutional filter,  F-filter represents frequency domain filter}
\centering
\begin{tabular}{lcccc}
\hline
                    & T-filter & F-filter & F-T filter & T-F filter (Ours) \\ \hline
PU-N09M07F10 (-4dB) & 63.01\%  & 56.48\%  & 58.23\%    & 92.08\%           \\
JNU (-10dB)         & 82.28\%  & 89.26\%  & 88.44\%    & 93.06\%           \\
HIT (-10dB)         & 61.99\%  & 60.83\%  & 64.11\%    & 86.15\%           \\ \hline
Average             & 69.09\%  & 68.86\%  & 70.26\%    & \textbf{90.43\%}           \\ \hline
\end{tabular}
\label{tab:comparefilter}
\end{table}

\section{Conclusions}

In this study, we have introduced a novel approach termed as ClassBD for bearing fault diagnosis under heavy noisy conditions. ClassBD is composed of cascaded time and frequency neural BD filters, succeeded by a deep learning classifier. Specifically, the time BD filter incorporates quadratic convolutional neural networks (QCNN), and we have mathematically proved its superior capability in extracting periodic impulse features in the time domain. The frequency BD filter includes a fully-connected linear filter, supplementing the frequency domain filter subsequent to the time filter. Furthermore, a deep learning classifier is directly integrated to empower classification capabilities. We have devised a physics-informed loss function composed of kurtosis, $l_2/l_4$ norm, and cross-entropy loss to facilitate the joint learning. This unified framework transforms traditional unsupervised BD into supervised learning, providing interpretability due to its retention of conventional BD operations. Finally, comprehensive experiments conducted on three public and private datasets reveal that ClassBD outperforms other state-of-the-art methods. ClassBD represents the first BD method that can be directly applied to classification and it exhibits a good noise resistance, portability, and interoperability. Therefore, ClassBD holds a great potential for further generalization on other difficult tasks such as cross-domain and small sample issues in future research.

\section*{Appendix}
The monotonicity of the sparsity objective function is derived as follows~\cite{he2021extracting, Liao_2023}:

First, given the input vector $\boldsymbol{x} \in \mathbb{R}^K$, the $G-l_p/l_q$ norm can be unfolded as:
\begin{equation}
\begin{aligned}
\mathcal{G}_{l_p/l_q}(\boldsymbol{x})&=\left( \frac{\left\| \boldsymbol{x} \right\| _{l_p}}{\left\| \boldsymbol{x} \right\| _{l_q}} \right) ^p
\\
&=\frac{\sum_{n=1}^K{|x\left( n \right) |^p}}{\left( \sum_{n=1}^K{|x\left( n \right) |^q} \right) ^{p/q}}
\\
&=\frac{\sum_{n=1,n\ne n^{\max}}^K{|x\left( n \right) |^p+  |x(n^{\max})|^{p}}}{\left( \sum_{n=1,n\ne n^{\max}}^K{|x\left( n \right) |^q+  |x(n^{\max})|^{q}} \right) ^{p/q}},
\end{aligned}
\end{equation}
where, $n^{\max}$ is the index of the maximum of $|\boldsymbol{x}|$. 

Such that, we calculate the derivative of $\mathcal{G}_{l_p/l_q}$ with respect to the maximum of $\boldsymbol{x}$:

\begin{equation}
\begin{aligned}
&\frac{\partial \mathcal{G}_{l_p/l_q}\left( \boldsymbol{x} \right)}{\partial \left( |x\left( n^{\max} \right) | \right)}=\\
&\frac{p|x\left( n^{\max} \right) |^{p-1}\left( \sum_{n=1,n\ne n^{\max}}^N{|x\left( n \right) |^q+|x\left( n^{\max} \right) |^{q}} \right) ^{p/q}}{\left( \sum_{n=1,n\ne n^{\max}}^N{|x\left( n \right) |^q+|x\left( n^{\max} \right) |^{q}} \right) ^{2p/q}}\\
&-\frac{\left( \frac{p}{q} \right) \left( \sum_{n=1,n\ne n^{\max}}^N{|x\left( n \right) |^q+|x\left( n^{\max} \right) |^{q}} \right) ^{p/q-1}q|x\left( n^{\max} \right) |^{q-1}\left( \sum_{n=1,n\ne n^{\max}}^N{|x\left( n \right) |^p+|x\left( n^{\max} \right) |^{p}} \right)}{\left( \sum_{n=1,n\ne n^{\max}}^N{|x\left( n \right) |^q+|x\left( n^{\max} \right) |^{q}} \right) ^{2p/q}}
\\
&=\frac{p|x\left( n^{\max} \right) |^{p-1}\left( \sum_{n=1,n\ne n^{\max}}^N{|x\left( n \right) |^q+|x\left( n^{\max} \right) |^{q}} \right) ^{p/q}}{\left( \sum_{n=1,n\ne n^{\max}}^N{|x\left( n \right) |^q+|x\left( n^{\max} \right) |^{q}} \right) ^{2p/q}}
\\
&-\frac{p|x\left( n^{\max} \right) |^{q-1}\left( \sum_{n=1,n\ne n^{\max}}^N{|x\left( n \right) |^q+|x\left( n^{\max} \right) |^{q}} \right) ^{p/q-1}\left( \sum_{n=1,n\ne n^{\max}}^N{|x\left( n \right) |^p+|x\left( n^{\max} \right) |^{p}} \right)}{\left( \sum_{n=1,n\ne n^{\max}}^N{|x\left( n \right) |^q+|x\left( n^{\max} \right) |^{q}} \right) ^{2p/q}}.
\end{aligned}
\label{eq:derivative}
\end{equation}



Then, the numerator of (\ref{eq:derivative}) is simplified to:
\begin{equation}
\begin{aligned}
&|x\left( n^{\max} \right) |^{p-1}\bigg( \sum_{n=1,n\ne n^{\max}}^N{|x\left( n \right) |^q+|x\left( n^{\max} \right) |^{q}} \bigg) 
-|x\left( n^{\max} \right) |^{q-1}\bigg( \sum_{n=1,n\ne n^{\max}}^N{|x\left( n \right) |^p+|x\left( n^{\max} \right) |^{p}} \bigg) 
\\
=&\frac{\left( \sum_{n=1,n\ne n^{\max}}^N{|x\left( n \right) |^q+|x\left( n^{\max} \right) |^{q}} \right)}{|x\left( n^{\max} \right) |^q} - \frac{\left( \sum_{n=1,n\ne n^{\max}}^N{|x\left( n \right) |^p+|x\left( n^{\max} \right) |^{p}} \right)}{|x\left( n^{\max} \right) |^p},
\end{aligned}
\end{equation}
when $p>q>0$, the following inequality holds:
\begin{equation}
\frac{\sum_{n=1,n\ne n^{\max}}^N{|x\left( n \right) |^q+ |x\left( n^{\max} \right) |^{q}}}{|x\left( n^{\max} \right) |^{q}}>\frac{\sum_{n=1,n\ne n^{\max}}^N{|x\left( n \right) |^p+|x\left( n^{\max} \right) |^{p}}}{|x\left( n^{\max} \right) |^{p}}.
\label{eq:ineq}
\end{equation}

Therefore, substituting (\ref{eq:ineq}) into (\ref{eq:derivative}), we have:
\begin{equation}
    \frac{\partial \mathcal{G}_{l_p/l_q}\left( \boldsymbol{x} \right)}{\partial|x\left( n^{\max} \right)|} > 0.
\label{eq:geq}
\end{equation}
Similarly, when $0 < p < q$, we have:
\begin{equation}
    \frac{\partial \mathcal{G}_{l_p/l_q}\left( \boldsymbol{x} \right)}{\partial |x\left( n^{\max} \right)|} < 0.
\label{eq:leq}
\end{equation}
Eq.~(\ref{eq:geq}) and Eq.~(\ref{eq:leq}) demonstrate that the $\mathcal{G}_{l_p/l_q}$ is a monotonic function with respect to the relationship between $p$ and $q$. Therefore, we set the loss function of $l_4/l_2$ norm and $l_2/l_4$ norm with the opposite sign to keep the optimization direction consistent.



\bibliographystyle{model1-num-names}

\bibliography{REF}





\end{document}